\documentclass[12pt,preprint,modern]{aastex62}

\usepackage{amssymb}
\usepackage{times}
\usepackage{verbatim}
\usepackage{color}
\usepackage{xspace}
\usepackage{lineno}

\usepackage{savesym}
\savesymbol{tablenum}
\usepackage{siunitx}
\restoresymbol{SIX}{tablenum}

\newcommand{\mone}  {^{-1}}
\newcommand{\mtwo}  {^{-2}}

\newcommand{\chan}    {{Chandra}\xspace}

\newcommand{\cmmtwo}  {\,\mathrm{cm\mtwo}}

\newcommand{\eflux}   {\,\mathrm{ergs\,cm\mtwo\,s\mone}}
\newcommand{\pflux}   {\,\mathrm{photons\,cm\mtwo\,s\mone}}
\newcommand{\elum}    {\,\mathrm{ergs\,s\mone}}
\newcounter{ion}     \newcommand{\eli}[2]  {\setcounter{ion}{#2}#1{~\sc\roman{ion}}}

\newcommand{\hetgs}   {{HETGS}\xspace}
\newcommand{\kev}     {\,\mathrm{keV}}
\newcommand{\kms}     {\,\mathrm{km\,s\mone}}

\newcommand{\mang}    {\,\mathrm{{\mbox{\AA}}}\xspace}
\newcommand{\mk}      {\,\mathrm{MK}}

\newcommand{\msun}  {\,M_\odot}
\newcommand{\kpc}   {\,\mathrm{kpc}}

\newcommand{\wrtf}  {{WR~25}\xspace}

\usepackage{csvsimple}

\usepackage{tikz,xcolor,hyperref}

\definecolor{lime}{HTML}{A6CE39}
\DeclareRobustCommand{\orcidicon}{
  \begin{tikzpicture}
    \draw[lime, fill=lime] (0,0) 
    circle [radius=0.16] 
    node[white] {{\fontfamily{qag}\selectfont \tiny ID}};
    \draw[white, fill=white] (-0.0625,0.095) 
    circle [radius=0.007];
  \end{tikzpicture}
  \hspace{-2mm}
}
\foreach \x in {A, ..., Z}{\expandafter\xdef\csname orcid\x\endcsname{\noexpand\href{https://orcid.org/\csname orcidauthor\x\endcsname}
    {\noexpand\orcidicon}}
}

\shorttitle{The Colliding Winds of \wrtf in High Resolution X-rays}

\begin{document}

\title{The Colliding Winds of \wrtf in High Resolution X-rays}

\author{\orcidA{}Pragati Pradhan}
\affiliation{Massachusetts Institute of Technology ,  77 Massachusetts Ave.,  Cambridge, MA 02139, USA}

\author{\orcidB{}David P.\ Huenemoerder}
\affiliation{Massachusetts Institute of Technology ,  77 Massachusetts Ave.,  Cambridge, MA 02139, USA}

\author{\orcidC{}Richard Ignace}
\affiliation{Department of Physics \& Astronomy,
  East Tennessee State University,
  Johnson City, TN 37614 USA}

\author{\orcidD{}A.M.T. Pollock}
\affiliation{Department of Physics and Astronomy, University of Sheffield, Hounsfield Road, Sheffield S3 7RH, UK}

\author{\orcidE{}Joy S.\ Nichols}
\affiliation{Harvard-Smithsonian Center for Astrophysics,
  60 Garden Street,
  Cambridge, MA 02138, USA
}

\begin{abstract}
  \wrtf is a colliding-wind binary star system comprised of a very
  massive O2.5If*/WN6 primary and an O-star secondary in a 208-day period
  eccentric orbit.  These hot stars have strong, highly-supersonic
  winds which interact to form a bright X-ray source from
  wind-collision-shocks whose conditions change with stellar
  separation.  Different views through the WR and O star winds are
  afforded with orbital phase as the stars move about their orbits, allowing
  for exploration of wind structure in ways not easy or even possible
  for single stars.  We have analyzed an on-axis Chandra/HETGS
  spectrum of \wrtf obtained shortly before periastron when the X-rays emanating from the system are the brightest.  From the on-axis observations, we constrain the line fluxes, centroids, and widths of various emission lines, including
  He-triplets of Si XIII and Mg XI.  We have also been able to include
  several serendipitous off-axis HETG spectra from the archive and
  study their flux variation with phase. This is the first report on
  high-resolution spectral studies of \wrtf in X-rays.

\end{abstract}

\keywords{stars: WR-type --- stars: massive --- stars: individual
  (\object{\wrtf}) --- X-rays: stars}

\section{Colliding Winds of Massive Stars}

Massive stars have strong stellar winds, accelerated to high
velocities by radiation pressure on millions of UV lines \citep{CAK}.
Instabilities lead to high temperature shocks embedded within the wind
capable of producing detectable X-ray emissions \citep{Lucy:White:1980,
  Feldmeier:Puls:Pauldrach:1997}.  While the X-ray emitting plasma is
thought to be a fairly small fraction of the total wind volume, X-ray
spectra provide key diagnostics of this plasma through the emission
lines and their profiles: line ratios are sensitive to the luminosity
and temperature distribution \citep[]{Kahn:Leutenegger:al:2001,
  Waldron:Cassinelli:2007, Walborn:Nichols:Waldron:2009}, line
profiles are sensitive to the wind structure \citep{ignace:2001,
  Owocki:Cohen:2001, Waldron:Cassinelli:2001, 2012MNRAS.420.1553S, 2016AdSpR..58..694I}.  Of fundamental
interest, particularly for X-ray spectral diagnostics, are the mass
loss rate ($\dot M$) and the momentum flux ($\dot M v_\infty$), with $v_\infty$ the
terminal velocity of the wind \citep[e.g.][]{2013MNRAS.429.3379O}.  Evolutionary models show that large
mass loss rates profoundly affect the ultimate state of the star while
the momentum input strongly affects the stellar environment
\citep[e.g.,][]{langer:2012, smith:2014}.  Their prodigious ionizing
UV radiation, of course, also affects their environments to large
distances.  This feedback can be an important part of galactic evolution.

O-stars, which have masses of $15$--$90\msun$, typically have terminal
velocities of about $2000\kms$ and mass loss rates of order
$\sim10^{-6}\msun\mathrm{yr^{-1}}$
\citep[e.g.,][]{puls:vink:al:2008}. Wolf-Rayet stars, which are at a
more advanced evolutionary stage typically having depleted H and
enriched in C, N, or O, have masses ranging from $10$--$100\msun$, and
terminal velocities of $\sim 2000\kms$, but much larger $\dot M
\gtrsim 10^{-5}\msun\mathrm{yr^{-1}}$
\citep{crowther:2007,hamann:grafener:al:2019}.

Some derived fundamental stellar parameters depend critically on model
assumptions regarding the degree of clumping in winds, or the wind
velocity with radius law, as examples, which may be difficult to
obtain empirically \citep[e.g., ][]{2011A&A...526A..32M, 2017A&A...603A..86S,  2018A&A...611A..17S, 2019A&A...631A.172D}.  A binary system provides the
opportunity to constrain some of these parameters.  Given two massive
stars, their winds collide at supersonic velocities resulting in
strong shocks \citep{1976AZh....53....6P, 1990ApJ...362..267L, 1992ApJ...389..635U, Stevens:Blondin:al:1992}.  At the relative velocities of a
few $1000\kms$, the shocks reach X-ray-emitting temperatures
($>10\mk$) and are also quite luminous, outshining the normal stellar
wind X-rays by an order of magnitude or more, as shown first for the
Wolf-Rayet stars including \wrtf\ by \citet{pollock1987}; this X-ray
emission becomes a primary diagnostic of the post-shock plasma \citep[e.g., ][]{2002A&A...383..636P, Henley:al:2003, 2010MNRAS.403.1657P}.  The
shock position and conditions depend upon the relative momentum flux
($\dot M v_\infty$) of each star and the stellar separations \citep[e.g., ][]{1996ApJ...469..729C, 2009ApJ...703...89G}.  In an
eccentric system, the stellar separation changes with orbital phase,
and we have the opportunity to probe different wind conditions, such
as the relative wind velocities at the stagnation point, different
orientations of the shocked region, and variations in the stellar radiation
fields in the stagnation region.  Line-of-sight absorption through cool pre-shock material also
occurs particularly near conjunction phases and can provide
clumping-independent estimates of the mass-loss rates \citep{2007ApJ...660L.141P}.

For models of colliding wind interactions, \citet{Stevens:Blondin:al:1992} presented the basic dynamics of wind
shocks with a 2D hydrodynamical analysis.  \citet{parkin:gosset:2011}
showed results of 3D hydrodynamical simulations of an eccentric
Wolf-Rayet system, including instabilities and generation of spiral
structure in the wind shock cone trailing the secondary.  Some classic examples of colliding wind binaries in eccentric orbits
are WR~140, with an 8-year period \citep{williams:al:1990,
  pollock:al:2005}, $\eta\,$Car in a $5.5\,$yr orbit
\citep{corcoran:al:2017,hamaguchi:al:2016}, or $\gamma^2\,$Vel in a 78
day orbit \citep{2001ApJ...558L.113S, cshild:al:2004}, to name but a few representative
studies which demonstrate the importance of orbital phase coverage in
X-rays of these types of systems.

In this paper, we will discuss one such colliding wind binary, \wrtf.
In section \ref{sec:wr25char}, we review previous X-ray studies of \wrtf\ along with current estimates for the binary properties. A description of the Chandra observations is presented in section
\ref{sec:obs}, with results from the high-resolution spectral analysis given in section
\ref{sec:highres}. Finally, line shapes, ratios, and overall
variability of \wrtf are described in section \ref{sec:disc}.

\section{Characteristics of \wrtf}
\label{sec:wr25char}

%
\begin{deluxetable}{ccl}
  \tablecaption{Properties of \wrtf}
  \tablehead{
    \colhead{Property}&
    \colhead{Value}&
    \colhead{Ref}
  }
  \startdata
  Spectral Type& 
  O2.5If*/WN6 + O&
  \citet{crowther:walborn:2011}
  \\
  &
  &
  \citet{Sota_2014}
  \\
  Minimum masses [$M_\odot$]& 
  $75\pm7$, $27\pm3$&
  \citet{gamen2008}
  \\
  Radii [$R_\odot$]&
  20.24, 13.52:&
  \citet{hamann:grafener:al:2019}
  \\
  &
  &
  \citet{howarth:vanleeuwen:2019}
  \\
  $v_\infty[\kms]$&
  2480, 2250:&
  \citet{hamann:grafener:al:2019}
  \\
  &
  &
  \citet{howarth:vanleeuwen:2019}
  \\
  $\log \dot M \mathrm{[M_\odot\,yr^{-1}]}$&
  $-4.6$, $-5.6$:&
  \citet{hamann:grafener:al:2019}
  \\
  &
  &
  \citet{howarth:vanleeuwen:2019}
  \\
  Period [days]&
  $207.85\pm0.02$&
  \citet{gamen:al:2006}
  \\
  HJD0&
  $2451598\pm1$&
  \citet{gamen:al:2006} 
  \\
  eccentricity&
  0.50:&
  \citet{gamen:al:2006}
  \\
  distance [kpc]& 
  $1.97^{+0.18}_{-0.15}$&
  \citet{rate:crowther:2020}
  \\
  %
  %
  \enddata
\end{deluxetable}\label{tbl:src}
%

The binary system \wrtf (HD 93162),
whose properties are noted in Table~\ref{tbl:src},
is comprised of an O2.5If*/WN6
\citep{crowther:walborn:2011} primary of minimum mass
$M_1=75\pm7\msun$ with an O-star secondary of minimum mass
$M_2=27\pm3\msun$ \citep{gamen2008} in an orbit with a 208 day period and an
eccentricity of about $e=0.50$ with an uncertainty of perhaps 0.10
\citep{gamen:al:2006}.
This makes the primary a contender for the most massive star in the
Milky Way, a status it shares with other
stars of similar spectral type of hydrogen-rich WN stars in binary
systems including
WR~20a, WN6ha+WN6ha \citep{rauw:debecker:al:2004}; 
WR~21a, O3/WN5ha+O3Vz((f*)) \citep{tramper:sana:al:2016}; and
WR~22, WN7h+O9III-V \citep{SSSSW:1999};
also all in Carina; and 
WR~43a, WN6ha+WN6ha \citep{schnurr:casoli:al:2008}
in the starburst-analog cluster NGC~3603.  Their binary periods range
from a few days to several months.  A similar set of perhaps even more
extreme stars is to be found in the lower-metallicity LMC including
Melnick~34 (Brey 84) which the Chandra T-ReX survey revealed as a
155-day period WN5h+WN5h binary of the highest X-ray luminosity
($>10^{35}\elum$) of any
colliding-wind binary including $\eta$~Carinae \citep{PCTBT:2018} and
the most massive known binary system with components of $139\msun$ and $127\msun$ \citep{TCBLPPS:2019}.
Evolutionary spectroscopic analysis shows that these stars are all
very young, very massive WR stars still burning hydrogen on the main
sequence in contrast to the more evolved classic WN and WC stars of
lower mass in which hydrogen is usually absent.

Apart from its minimum mass, little is known about the otherwise unclassified
O-star secondary although the mass does suggest that the well-studied O4If star
$\zeta$~Puppis \citep{howarth:vanleeuwen:2019} is a suitable analog for suggesting
the plausible set of secondary properties shown in Table~\ref{tbl:src}.

\begin{figure}[ht!]
  \centering\leavevmode
  \includegraphics*[width=0.75\columnwidth]{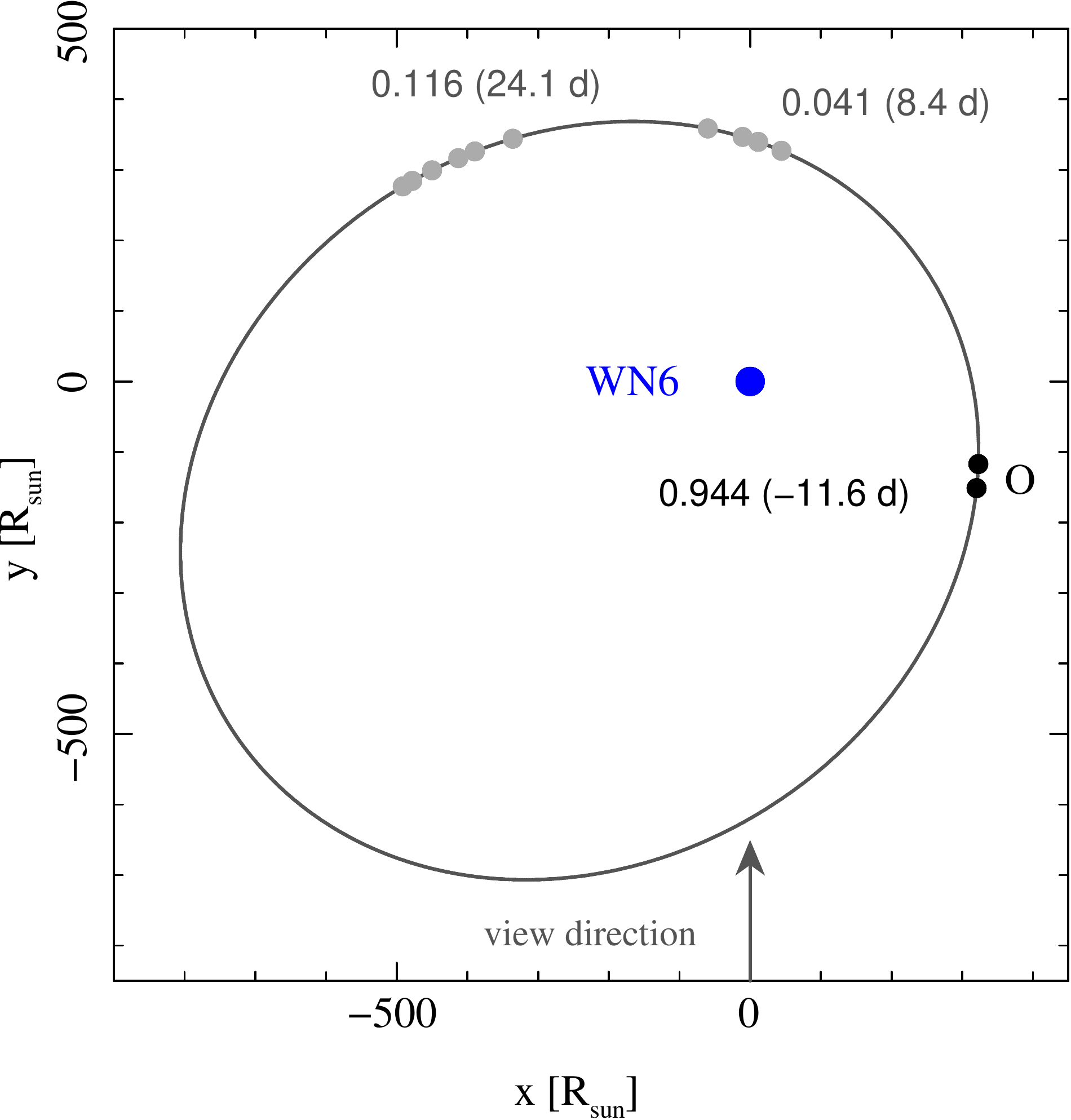}
  \caption{Schematic of the orbit of \wrtf, viewed from above the orbital plane,
    referenced to the position of the WR-star primary, shown as the
    blue circle. Relative positions of the
    O-star secondary are shown as small circles for the
    on-axis HETG observations in black, and for the
    serendipitous off-axis observations in gray.  Phases for the start of each group of observations
    are shown near the O-star circles, with days from periastron in
    parentheses.  }  \label{fig:orb}
\end{figure}

\begin{figure}[ht!]
  \centering\leavevmode
  \includegraphics*[width=0.90\columnwidth]{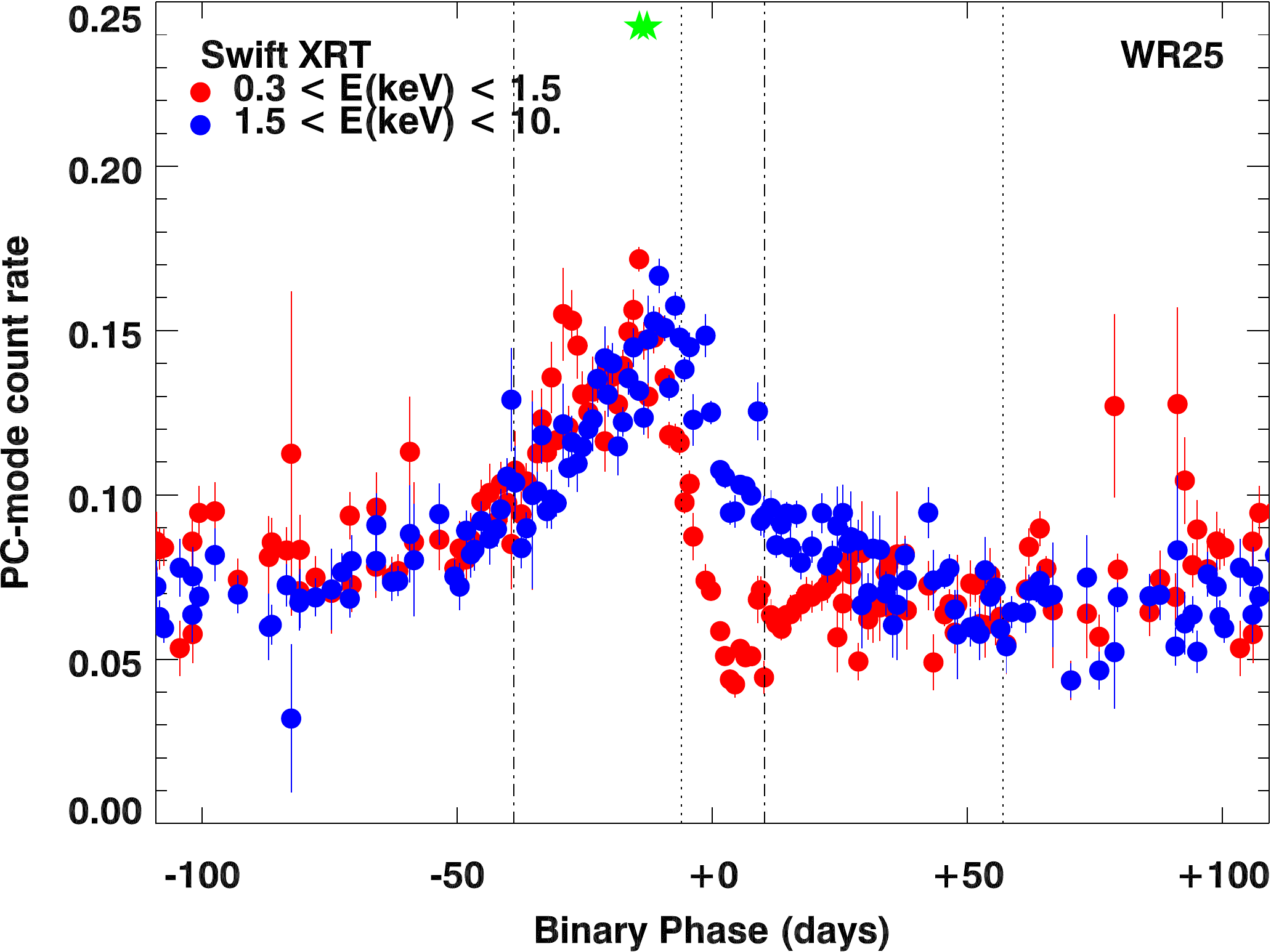}
  \caption{The {\it
      Swift} XRT light curve of \wrtf\ in two bands.  The soft band
    ($0.3$--$1.5\kev$) is shown in red, and the hard band
    ($1.5$--$10\kev$) in blue.  From the optical radial-velocity orbital
    solution, the vertical dotted lines mark quadratures (extrema in
    radial velocity), while the dot-dash vertical lines denote
    conjunctions, with the O-star in front about $40\,$days before
    periastron, and the WR-star in front at about $10\,$days after,
    where phase $0.0\,$days is periastron.  Two overlapping green stars
    mark the phases of the pointed \chan/HETGS observations at maximum light
    near quadrature corresponding to the black circles in Fig.~\ref{fig:orb}.}
  \label{fig:swift}
\end{figure}

\wrtf is the nearest and X-ray brightest of this important class
of very massive stars.  It
lies in the Carina region at a Gaia DR2 distance of $1.97 \kpc$
and has figured prominently in massive star studies since the
beginning of hot-star X-ray astronomy and the report of its discovery
in the first-light issue of the {\it Einstein} Observatory
\citep{seward1979}.  It is one of the very few colliding-wind binary
systems with complete orbital coverage. This has been procured through 266
separate {\it Swift} observations, mostly granted to one of us
through 13 separate ToO
requests for \wrtf\ under Target ID 31097 between 2007 and 2018, that have been
complemented with occasional more precise {\it XMM-Newton}, {\it
  Suzaku} and {\it NuSTAR} measurements, often available
serendipitously through observations of its neighbor $\eta$~Carinae.
The X-ray light curve shows the distinct X-ray brightening expected
from colliding winds moving towards periastron
\citep{pollock:2012, pandey:al:2014, arora:al:2019}.  In Figure~\ref{fig:orb} we show
the orbital geometry and in Figure~\ref{fig:swift} details of the
notably asymmetric phased light curve from {\it Swift}.  The
soft-to-hard flux ratio is constant throughout most of the orbit but
starts to decrease about 10 days before periastron as the WR primary
moves in front, reaching a well-defined minimum
about 2 days before the conjunction predicted by
the optical radial-velocity solution, which might not precisely track
the primary's orbital motion \citep{grant:blundell:al:2020} but may
also suggest a slightly higher orbital eccentricity.

From the mass-loss rates and terminal velocities in Table~\ref{tbl:src}
the stagnation point of the colliding flows is expected to be closer to the O-star, and thus
the shock cone convex towards the WR-star.  Analysis by
\citet{arora:al:2019} suggested that the wind velocity of each star at
the stagnation point changes significantly over the course of the
orbit, from about $1600\kms$ at periastron to $2200\kms$ at apastron,
although this does not coincide with any obvious softening of the
spectrum \citep{pollock2006}.  The shock may transition from adiabatic
to radiative near periastron, and perhaps even undergo radiative
braking \citep{gayley:al:1997} through proximity to the intense O-star
radiation field.  The change in wind-wind collision pre-shock relative
velocity from about $3200\kms$ to $4400\kms$ should result in a
corresponding change in the post-shock temperature more than $50$\%.

\begin{figure}[ht!]
  \centering\leavevmode
  \includegraphics*[width=0.85\columnwidth]{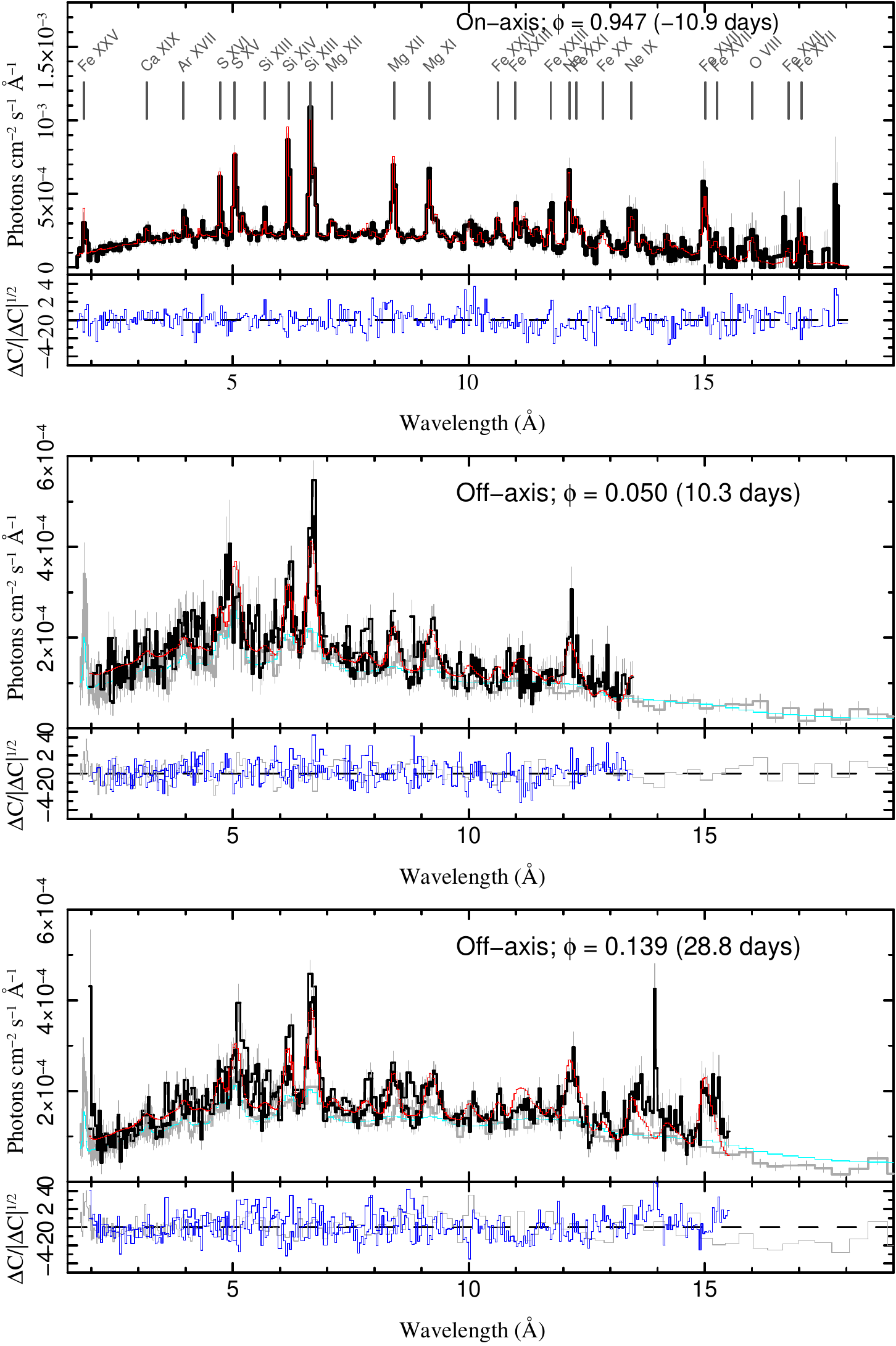} \caption{Top:
  Pointed \chan/HETGS \wrtf flux spectrum (black), a powerlaw emission
  measure plasma model fit (red) and residuals (blue, lower panel).
  Center and bottom plots show the serendipitous off-axis HETGS
  spectra with their fitted high-resolution models convolved with a Gaussian
  appropriate for the off-axis blurring. 
  The gray curve (with errorbars) is the zeroth order spectrum, 
  and its corresponding model is the light solid line.   The numbers in parentheses
  gives the days from periastron.  }  \label{fig:xrayspec}
\end{figure} 

\section{Observations}\label{sec:obs}
To better quantify the plasma conditions, we have obtained
Chandra/HETG high-resolution X-ray spectra slightly before periastron
(black circles in Figure~\ref{fig:orb}, green stars in  Figure~\ref{fig:swift}), at phases in 2016 August when the Swift rates predicted a maximum.  With these
observations we can potentially determine line centroids, resolve
lines widths, determine plasma temperatures from line strengths and
the continuum shape, and constrain X-ray source locations from the
photo-excitation in He-like ions' forbidden-to-intercombination line
ratios. From line-to-continuum ratios, we can explore possibilities of
unusual abundances or collisionless plasma effects.

Although quite far off-axis ($\sim 10.5\,\mathrm{arcmin}$), we have
also analyzed several serendipitous Chandra/HETG spectra of \wrtf
taken more than 10 years earlier during high-resolution observations
of other hot stars in Carina.
The instrumental profile at the source locations is however too broad to
characterize line widths or centroids accurately.  The
off-axis data provide some line and broad-band fluxes, useful for
study of the stellar wind and wind-shock region structure through variations with phase.

Table~\ref{tbl:obs} provides a log of the observations, exposure and
the orbital phases, both for the on-axis and off-axis pointings.

%
\begin{deluxetable}{rrccc}
  \tablecaption{Table of observations analyzed in the current work
    with relevant time of observation, ObsID, exposure and orbital
    phases } \tablehead{ \colhead{Start Time}& \colhead{ObsID}&
    \colhead{Exposure}& \colhead{$\phi$}& \colhead{$\Delta\phi$}\\
    &&[ks]&&[days]
  }
  \startdata
  \multicolumn{5}{c}{(on-axis pointings)}\\
  2016-08-10T15:19:33& \dataset[18616]{ https://heasarc.gsfc.nasa.gov/FTP/chandra/data/science/ao17/cat2//18616/}& 28& 0.944& -11.6\\
  2016-08-11T23:28:52& \dataset[19687]{https://heasarc.gsfc.nasa.gov/FTP/chandra/data/science/ao17/cat2//19687/}& 57& 0.951& -10.2\\\hline
  \multicolumn{5}{c}{(serendipitous off-axis pointings)}\\
  2005-11-08T07:19:58& \dataset[7201]{https://heasarc.gsfc.nasa.gov/FTP/chandra/data/science/ao06/cat2//7201/}& 18& 0.041& 8.4\\
  2005-11-09T14:44:53& \dataset[7204]{https://heasarc.gsfc.nasa.gov/FTP/chandra/data/science/ao06/cat2//7204/}& 33& 0.047& 9.7\\
  2005-11-10T11:02:37& \dataset[7202]{https://heasarc.gsfc.nasa.gov/FTP/chandra/data/science/ao06/cat2//7202/}& 10& 0.051& 10.6\\
  2005-11-12T07:41:15& \dataset[7203]{https://heasarc.gsfc.nasa.gov/FTP/chandra/data/science/ao06/cat2//7203/}& 27& 0.060& 12.4\\
  2005-11-29T22:35:38& \dataset[5397]{https://heasarc.gsfc.nasa.gov/FTP/chandra/data/science/ao06/cat2//5397/}& 10& 0.145& 30.1\\
  2005-12-01T12:19:53& \dataset[7228]{https://heasarc.gsfc.nasa.gov/FTP/chandra/data/science/ao06/cat2//7228/}& 20& 0.152& 31.6\\
  2005-12-02T08:44:16& \dataset[7229]{https://heasarc.gsfc.nasa.gov/FTP/chandra/data/science/ao06/cat2//7229/}& 20& 0.156& 32.5\\
  2006-06-19T18:00:43& \dataset[7341]{https://heasarc.gsfc.nasa.gov/FTP/chandra/data/science/ao06/cat2//7341/}& 53& 0.116& 24.1\\
  2006-06-22T10:18:52& \dataset[7342]{https://heasarc.gsfc.nasa.gov/FTP/chandra/data/science/ao06/cat2//7342/}& 49& 0.129& 26.7\\
  2006-06-23T17:45:18& \dataset[7189]{ https://heasarc.gsfc.nasa.gov/FTP/chandra/data/science/ao06/cat2//7189/}& 18& 0.135& 28.0\\
  \enddata
\end{deluxetable}\label{tbl:obs}
%

\section{High Resolution Spectral Analysis}\label{sec:highres}
\subsection{On-axis observations}\label{sec:onax}

The counts spectra for the observations were extracted using
\texttt{TGCat} reprocessing scripts \citep{Huenemoerder2011}, using
CIAO \citep{CIAO:2006} version 4.8 and Calibration Database version
4.7.2. Since the two on-axis observations are taken less than a day apart
with orbital
phases that differ by only $0.007$ (see Table \ref{tbl:obs}), we
combined the negative and positive first order spectra from both the
observations to get one combined spectrum each for HEG and MEG. The
advantage of doing this is to increase the signal-to-noise ratio
facilitating high resolution spectral analysis, and also reducing the
number of model bins, thereby reducing execution time in model
evaluation during fitting.

We fit the spectra with both a global plasma model, and line-by-line.
The global model gives a broad characterization of plasma conditions,
within the parameters of the model,  under the assumption that all
the lines have the same shape and Doppler offset.  The line-by-line
fits relax the latter assumptions, and are independent of any global
plasma model since they are local parametric functional fits.

For global spectral modeling, we fit the 1.5-20 \SI{}{\angstrom}
spectrum of \wrtf with a powerlaw emission measure distribution plasma
model, as described in detail by \citet{Huenemoerder:al:2020}. The
model is of the form,
$dEM / dT = [n_e n_H \, dV/dT] \propto (T/T_\mathrm{max})^{-\beta}$,
where $dEM/dT$ is the differential emission measure, $V$ is the volume
of X-ray-emitting plasma, $n_e$ and $n_H$ are the electron and
hydrogen number densities, respectively, $T$ is the plasma temperature
which we limit to a range $T_\mathrm{min}$, $T_\mathrm{max}$, and the
exponent on temperature is $\beta$.  This model is obtained from line
and continuum emissivites from the Astrophysical Plasma Emission Code
(APEC) as stored in the atomic database, AtomDB
\citep{smith2001,foster2012}.  We used the abundances from
\citet{Asplund:Grevesse:al:2009}.  \wrtf likely has a reduced abundance
of hydrogen and increased helium in the WN component.  Hence
interpretation of the normalization and relative abundances requires
caution; \citet{2000MNRAS.318..214I} and \citet{schulz:al:2019pp} provide a method for scaling of
parameters in low hydrogen plasmas.  The model also assumes that all
emission lines have the same Gaussian shape and Doppler shift, and
that neutral absorption is all in the foreground.  As such, the model
represents the mean characteristics of the emitting plasma, and is not
a detailed dynamical and geometrical model.  Uncertainties on model
parameters were determined from a Markov-Chain, Monte-Carlo
evaluation.  \citep[an ISIS implementation, ``{\tt emcee}'', of the
][Python code]{emcee:2013} to search parameter space and form
confidence contours.\footnote{The ISIS version is described at
  \url{https://www.sternwarte.uni-erlangen.de/wiki/index.php/Emcee}
  and is available as part of the {\tt isisscripts} package available
  at \url{https://www.sternwarte.uni-erlangen.de/isis/}. }

The data, fitted model, and residuals are shown in the top panel of
Fig.~\ref{fig:xrayspec}. The spectrum of \wrtf has many strong
emission lines, primarily of H- and He-like ions, but also of many Fe
ions. Below about $10\mang$, the underlying continuum is also
prominent.  Model parameters are given in
Table~\ref{tbl:plawdemparams}.
%

\begin{deluxetable}{c|cc|cc|cc|c}
  \tablecaption{Emission Measure Model Parameters}
  \tablehead{
    \colhead{Parameter}&
    \colhead{Value}&
    \colhead{$\sigma$}&
    \colhead{Value}&
    \colhead{$\sigma$}&
    \colhead{Value}&
    \colhead{$\sigma$}&
    \colhead{Units}\\
    & \multicolumn{2}{c|}{$\phi=0.95$}&  \multicolumn{2}{c|}{$\phi=0.05$}& \multicolumn{2}{c|}{$\phi=0.13$}& 
  }
  \startdata
  $norm$&                $0.032$&     $0.003$&      $0.025$& 0.003& $0.024$&0.002&    $\mathrm{cm^{-5}}$\\
  $\beta$&               $1.02$&      $0.08$&       $1.22$& 0.09&  $1.18$& 0.07&    \\
  $\log T_\mathrm{max}$& $7.70$&      $0.03$&       $7.93$& 0.04&  $7.82$& 0.03&    dex K\\
  $\log T_\mathrm{min}$& $6.00$&      (frozen)&     --&--&--&--&      dex K\\
  $v_\mathrm{turb}$&     $1338$&      $42$&         --&--&--&--&      $\kms$\\
  $v_\mathrm{Doppler}$&  $-303$&      $32$&         --&--&--&--&      $\kms$\\
  $A(\mathrm{Ne})$&      $0.60$&      $0.09$&       $0.48$& $0.11$& $0.42$& $0.05$& \\
  $A(\mathrm{Mg})$&      $0.64$&      $0.04$&       $0.51$& $0.06$& $0.41$& $0.04$& \\
  $A(\mathrm{Si})$&      $0.85$&      $0.05$&       $1.02$& $0.08$& $0.81$& $0.05$& \\
  $A(\mathrm{S})$&       $1.10$&      $0.10$&       $1.73$& $0.20$& $1.20$& $0.10$& \\
  $A(\mathrm{Fe})$&      $0.48$&      $0.05$&       $0.37$& $0.06$& $0.56$& $0.03$&\\
  $N_\mathrm{H}$&        $0.65$&      $0.04$&       $0.74$& $0.05$& $0.56$& $0.03$&  $10^{22}\,\mathrm{cm}^{-2}$\\\hline
  $f_x(obs)$& \multicolumn{2}{c|}{\num{1.01e-11}}& \multicolumn{2}{c|}{\num{6.14e-12}} &\multicolumn{2}{c|}{\num{6.11e-12}}&   $\eflux$\\
  $f_x(N_\mathrm{H}=0)$& \multicolumn{2}{c|}{\num{3.70e-11}}& \multicolumn{2}{c|}{\num{2.98e-11}} &\multicolumn{2}{c|}{\num{2.71e-11}}&  $\eflux$\\
  \enddata
  \tablecomments{The normalization is related to the emission
    measure via $norm = [10^{-14}/(4\pi\,d^2)] \int{n_\mathrm{e} n_\mathrm{H}\,
      dV}$, subject to re-interpretation for H-deficient, He-rich
    plasmas as stated in the text.  The broadening was treated as a
    ``turbulent'' velocity term, added in quadrature to the thermal
    velocity and common to all ions.  Abundances, $A(Z)$, are
    fractions by number relative to \citet{Asplund:Grevesse:al:2009}
    (and were 1.0 if not listed).  Uncertainties 
    ($\sigma$) are given as a 68\%
    confidence error-bars, as determined with Markov-Chain,
    Monte-Carlo methods.  For the off-axis data ($\phi =
    0.05,\,0.13$), unspecified parameters were taken from the
    on-axis fits and frozen.  Fluxes were evaluated from the fitted
    model over the band $1.0$--$40\mang$ ($0.3$--$12\kev$). 
    \citet{pandey:al:2014} estimated that the non-local ISM column to be about $0.37\times10^{22}\cmmtwo$.
  }
\end{deluxetable}\label{tbl:plawdemparams}



Parametric line fitting was performed using the plasma model as a
guide to line identification and probable blend assessment.  Emission
lines were fit in groups of overlapping or close features, using sums
of Gaussian profiles folded through the instrument response. If
features were too blended or too weak for unconstrained fits, they
were restricted accordingly, either by constraining its 
parameters to a stronger
line's parameters (such as in using wavelength offsets or tied
Gaussian $\sigma$), or by freezing the position or width and fitting
the flux.  The plasma model also served to provide a continuum model,
since in crowded regions, with wind broadening, the local minima are
not good approximations to the continuum.  In these line fit models,
we did not include neutral absorption, hence, the flux is as-observed
at Earth.  Emission line fit results are given in Table
\ref{tbl:lineparams}; for tied parameters in constrained fits,
uncertainties have null values. The regions fit are shown in
Figures~\ref{fig:lineprof1}, \ref{fig:lineprof2}, \ref{fig:lineprof3}.

We also examined the line profiles of strong lines in detail, considering whether a
Gaussian shape is adequate, and we studied in detail the He-like
ratios for evidence of photoexcitation in the
forbidden-to-intercombination line ratio. Results will be discussed
in subsequent sections.

\begin{deluxetable}{cr|cc|cc|cc}
  \tablecaption{Emission Line Parameters for the on-axis observations.}
  \tablehead{
    \colhead{Line} & 
    \colhead{$\lambda_0$} &  
    \colhead{$\Delta v$} &  
    \colhead{$\sigma_{\Delta v}$}  &   
    \colhead{$f_x$}& 
    \colhead{$\sigma_{f}$}&  
    \colhead{$FWHM$}  & 
    \colhead{$\sigma_{FW}$} \\
    \hline 
    & 
    \multicolumn{1}{c|}{\AA}& 
    \multicolumn{2}{c|}{$\kms$}   &
    \multicolumn{2}{c|}{\footnotesize{$10^{-5}\pflux$}}&
    \multicolumn{2}{c}{$\kms$}
  }
\startdata
Fe XXV&
1.858&
--&
--&
1.64&
0.38&
--&
--
\\
Ar XVII (r+i) &
3.949 &
-56&
238&
0.69&
0.18&
1595&
497
\\
Ar XVII (f)&
3.994 &
--&
--&
0.62&
0.17&
--&
--
\\
S XVI&
4.730 &
162&
215&
2.55&
0.35&
2915&
476
\\
S XV (r)&
5.039&
-278&
310&
2.62&
0.68&
2453&
446
\\
S XV (i)&
5.065&
 --&
-- &
0.84&
0.61&
--&
--
\\
S XV (f)&
5.102&
 --&
-- &
1.44&
0.34&
--&
--
\\
Si XIII $\beta$&
5.681&
 -266&
326&
1.18&
0.28&
2575&
731
\\
Si XIV&
6.183&
-263&
65&
4.83&
0.26&
2339&
149
\\
Si XIII (r)&
6.648&
 -242&
67&
5.49&
0.37&
2303&
135
\\
Si XIII (i)&
6.687&
 --&
-- &
0.96&
0.33&
--&
--
\\
Si XIII (f)&
6.740&
--&
-- &
2.97&
0.23&
--&
--
\\
Mg XII&
8.422&
-332&
80&
3.73&
0.31&
2160&
198
\\
Mg XI (r)&
9.169&
 -291&
133&
3.20&
0.37&
1985&
275
\\
Mg XI (i)&
9.230&
 --&
-- &
0.57&
0.36&
--&
--
\\
Ne X&
9.291&
 --&
-- &
0.52&
0.51&
--&
--
\\
Mg XI (f)&
9.314&
--&
-- &
0.47&
0.52&
--&
--
\\
Ne X&
9.362&
 --&
-- &
0.38&
0.29&
--&
--
\\
Ne X&
9.481&
 --&
-- &
0.20&
0.21&
--&
--
\\
Ne X&
9.708&
--&
--&
0.26&
0.20&
--&
--
\\
Ne X&
12.135&
 -387&
128&
6.33&
0.59&
2568&
292
\\
Fe XVII &
12.266&
--&
-- &
2.74&
0.47&
--&
--
\\
Fe XXI&
12.390&
--&
--&
2.78&
0.48&
--&
 --
\\
Ne IX (r)&
13.447&
-526&
147&
2.99&
0.68&
1695&
246
\\
Fe XIX&
13.497&
--&
-- &
0.11&
0.48&
--&
--
\\
Ne IX (i) &
13.552&
 --&
-- &
2.77&
0.70&
--&
--
\\
Ne IX (f) &
13.699&
--&
-- &
1.69&
0.51&
--&
--
\\
Fe XVII&
13.825&
 --&
-- &
0.94&
0.45&
--&
--
\\
Fe XVII&
15.014&
-477&
183&
5.77&
1.10&
2466&
498
\\
Fe XVII&
15.261&
--&
--&
1.02&
0.52&
--&
--
\\
O VIII&
16.006&
-561&
184&
3.56&
1.24&
3441&
1781
\\
Fe XVII&
17.051&
--&
--&
2.04&
1.14&
--&
--
\\
Fe XVII&
17.096&
 --&
--&
1.24&
0.87&
--&
--
\\
\enddata
\tablecomments{
 $\lambda_0$ is the theoretical wavelength and 
   for each line, the velocity, error on velocity, flux, error on flux and full-width half maximum ($FWHM$), error on $FWHM$ are denoted by $\Delta v$, $\sigma_{\Delta v}$, $f_x$, $\sigma_f$, $FWHM$ and $\sigma_{FW}$ respectively. Only fluxes are reported for the  lines where Gaussian width was frozen and where the wavelength was tied to their relative offsets from the resonance line. All errors are calculated for $1$-$\sigma$ confidence level limits.
}
\end{deluxetable}\label{tbl:lineparams}



\subsection{Serendipitous Off-axis Observations}\label{sec:offaxisobs}

We  analyzed 10 off-axis serendipitous observations of \wrtf (gray circles in Fig.~\ref{fig:orb} at phases $\sim0.05$ and $\sim0.14$).  These were taken with the source at about $10\,\mathrm{arcmin}$ off-axis which significantly blurs the spectrum due to the large point-spread-function.  The spatially broad spectra (in the cross-dispersion direction) were also truncated obliquely in some orders by the detector boundary, making the calibration uncertain.  Hence we ignored wavelengths beyond about $14\mang$.  The shortest wavelengths, below about $2\mang$, were also compromised by overlap of the wide HEG and MEG spectral arms, precluding extraction of \eli{Fe}{25}.

To fit the HEG/MEG spectra, we determined the blur scale from the zeroth order extent, then used a Gaussian convolution model (Xspec's ``{\tt gsmooth}'') on our APEC-based powerlaw emission measure model. Since the wavelength range of the dispersed off-axis spectrum is somewhat limited, we also simultaneously fit the zeroth-order spectra.  Given the off-axis blurring of the image, CCD pile-up was not an issue, as it is for the on-axis zeroth order.  To account for possible systematic differences in the zeroth order and dispersed effective area calibrations, we introduced an arbitrary renormalization factor on the zeroth order, whose value was $0.80\pm0.01$, which is somewhat larger than the nominal $5\%$ expected calibration uncertainty. We fit this model to spectra in two phase groups (freezing the broadening and Doppler shift to the high-resolution fit values).  Uncertainties were determined via {\tt emcee}.
The results are shown in Figure~\ref{fig:xrayspec} and fitted parameters are listed in Table~\ref{tbl:plawdemparams}.

For the zeroth order spectra, we performed analysis only for the off-axis observations, since the on-axis spectra were heavily piled-up. We downloaded the relevant ObsIDs from \texttt{TGCat} \citep{Huenemoerder2011} and used \texttt{dmextract} to extract the source spectrum by selecting circular regions of 60 arc-seconds centered on the source position. The relevant response files and the ancillary files were generated with \texttt{mkacisrmf} and \texttt{mkarf}. We corrected the ancillary files for the aperture efficiency factor using \texttt{arfcorr} as recommended in the CIAO website\footnote{https://cxc.cfa.harvard.edu/ciao/ahelp/}.

All spectral fitting was performed using the Interactive Spectral Interpretation System \citep[ISIS software\footnote{\url{https://space.mit.edu/cxc/ISIS/index.html}},][]{houck2000}, which also provides interfaces to AtomDB as well as to {\tt xspec} \citep{Arnaud:1996} models. 

\clearpage
\section{Discussion}\label{sec:disc}

In a colliding-wind binary, the shocked-wind region far outshines the
X-ray emission from the components' stellar winds in isolation.  The
components of the \wrtf system are similar to $\zeta\,$Pup and WR~6,
which have luminosities of about $6\times10^{29}\elum$ and
$8\times10^{32}\elum$, respectively.  If we assume the interstellar
line-of-sight column density to \wrtf of $3.7\times10^{21}\cmmtwo$
estimated by
\citet{pandey:al:2014}, then the luminosity implied by the spectrum at
$\phi=0.96$ is about $9\times10^{33}\elum$, nearly an order of
magnitude larger than that expected from the individual stellar winds in
isolation.  The changing distance between the components around an
eccentric orbit causes the wind-wind collision to occur at different
velocities in the accelerating matter, and the varying aspect lets us
view the shock from different angles, as well as  through
different wind columns of the two stars.  Using high-resolution
spectroscopy, we can therefore study the temperature and dynamics of
the shock through line fluxes and profiles.  The broad-band spectral
shape is also sensitive to temperature and the line-of-sight neutral
absorption.

Ideally, we would like to monitor the emission lines around the orbit.
Hence, given only high-resolution spectra at just before periastron,
this study represents only a start at a detailed view of \wrtf in
high-resolution X-rays.

\begin{figure}
  \centering\leavevmode
  \includegraphics[width=0.6\columnwidth]
{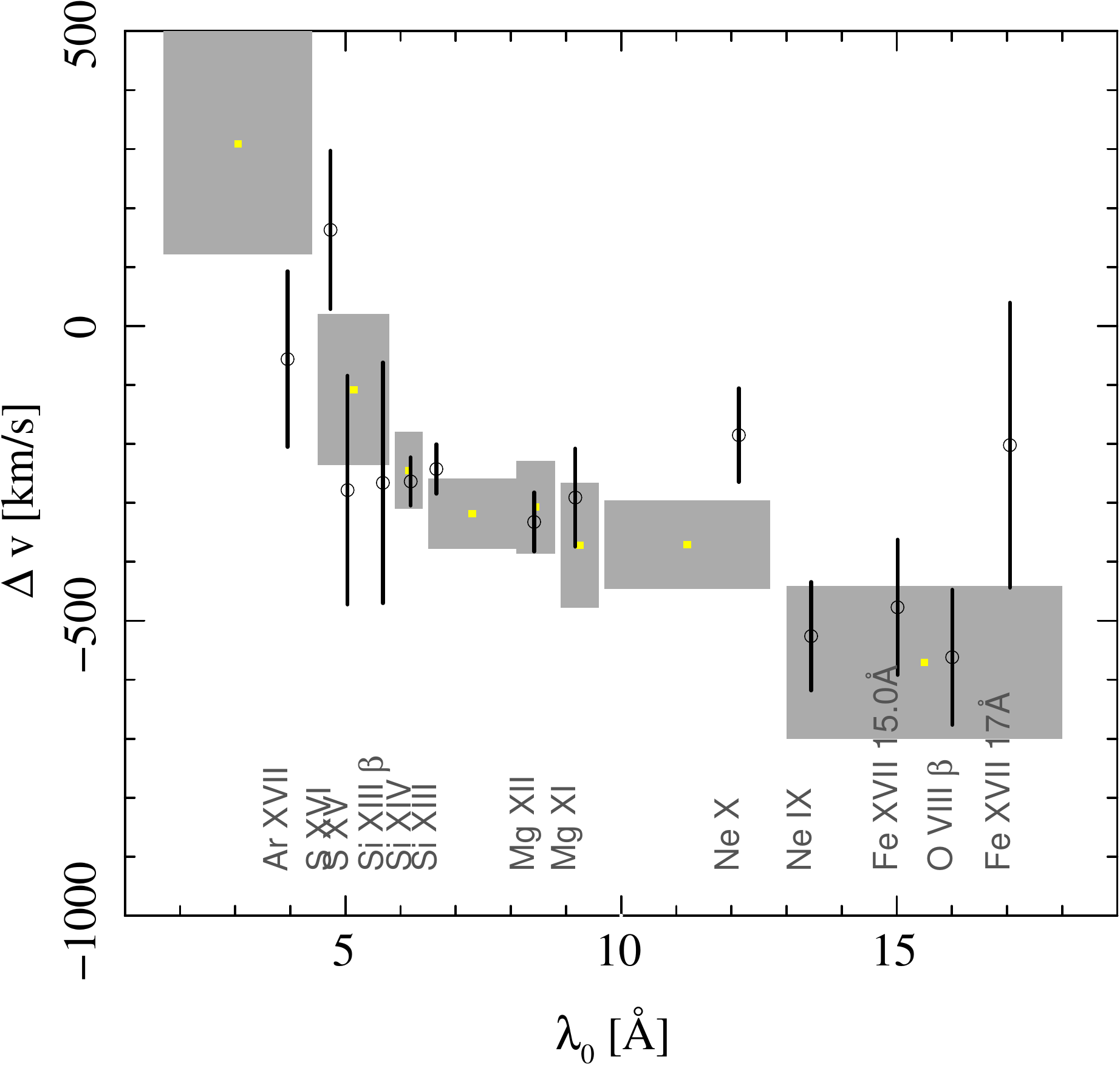}
\caption{The line centroid velocity offsets from rest wavelengths is
  shown for two different approaches.  The black circles with
  errorbars show the values from the Gaussian fits to line groups.
  The gray shaded rectangles show the result of piece-wise plasma
  model fits for the Doppler shift and broadening in several
  wavelength regions.  Errorbars and gray shading vertical extent
  indicate one standard deviation uncertainties, while the gray
  rectangles' widths indicate the wavelengthr region fit. }
  \label{fig:dvwav}
\end{figure}

From the local line fits to the on-axis observations near
$\phi = 0.94$ (see Table~\ref{tbl:lineparams} and Figures~\ref{fig:lineprof1}, \ref{fig:lineprof2}, and \ref{fig:lineprof3}), we find that there is a general blueshift
of the centroid from the rest wavelength by about $-300\kms$, with a
possible decreasing trend in centriod with increasing wavelength.  In
Figure~\ref{fig:dvwav}, the points with errorbars show these line-fit
results.  We also explored another approach, using the global plasma model
as a starting point, and re-fit only the normalization, Doppler shift,
and broadening in several wavelength regions.  The global model, in
contrast, used a common offset and broadening for all lines.  This is
of course not independent from the Gaussian line fitting, but
takes advantage of the multiplexing of many features in the region, and
using the underlying plasma model implicitly handles line blends.
The result is shown as the shaded regions in Figure~\ref{fig:dvwav},
and the trend with wavelength is very clear.  

\begin{figure}
  \centering\leavevmode
  \includegraphics[width=0.6\columnwidth]{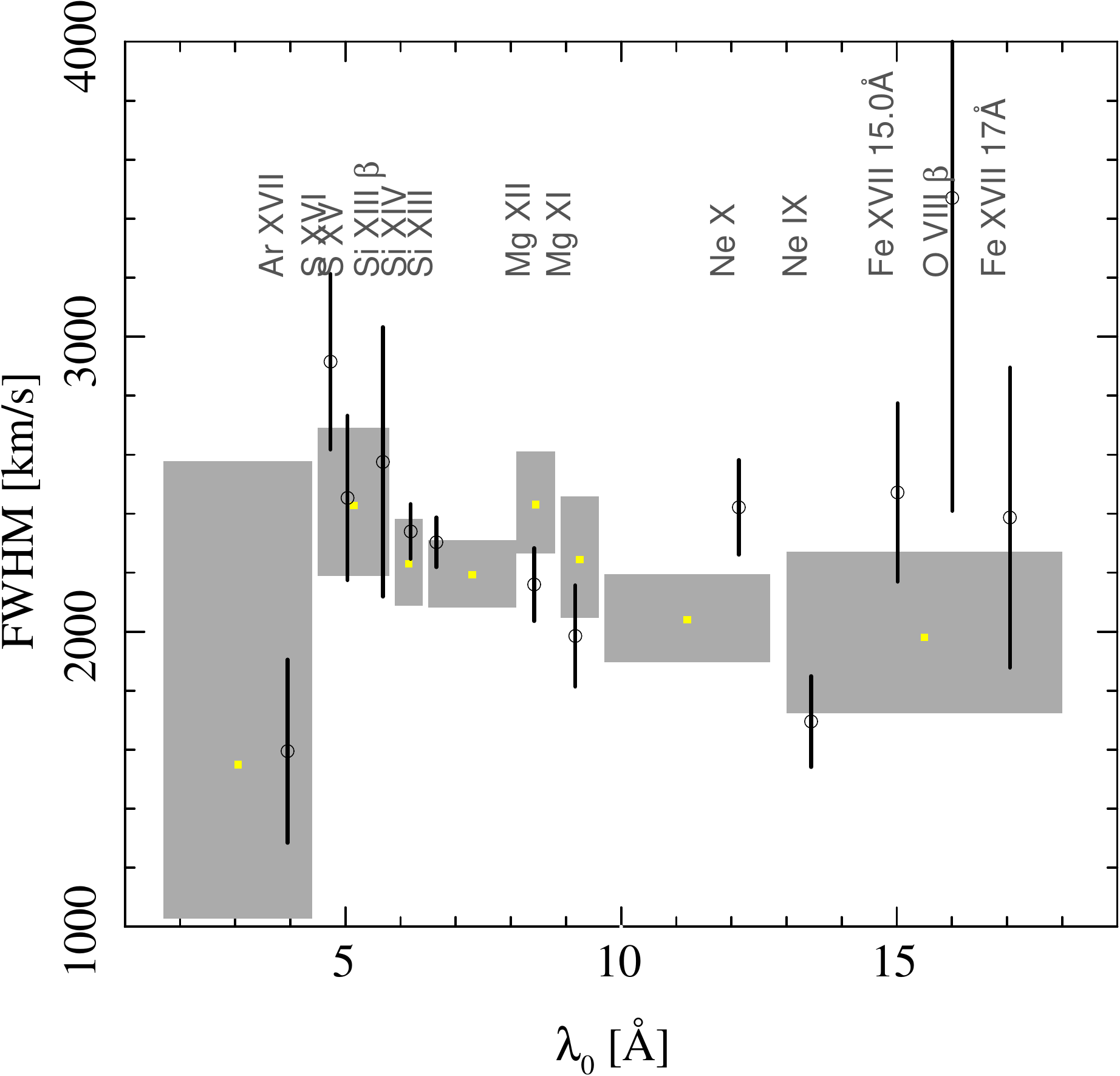} \caption{The
  line widths are shown from the Gaussian fits to line
   groups as points with errorbars, and from piece-wise plasma model
   fits, as in Figure~\ref{fig:dvwav}. }
  \label{fig:fwhmwav}
\end{figure}

Since we are viewing the wind-wind collision shock-cone nearly
edge on (perpendicularly to the line between stellar centers), 
we should expect broad lines with an apparent blue-shift
if the receding material is obscured by foreground absorption by the
shock-cone flow or stellar wind neutral plasma.  In
Figure~\ref{fig:fwhmwav}, we can see that the lines are broad, and
seem independent of wavelength.

The ratio of wind momenta,
$\eta = ( \dot M_s v_{\infty,s} ) / ( \dot M_p v_{\infty,p} )$, is in
the range of $0.2$ to $0.5$, given the parameters in
\citet{arora:al:2019}.  For these values, \citet{pittard:dawson:2018}
find opening angles of the primary (WR-star) shock to be roughly
$60^\circ$--$90^\circ$ (from the secondary to the stagnation point to
the contact discontinuity line or primary shock envelope).  At
$\phi=0.94$ we should have a significant amount of plasma moving parallel to the
line-of-sight \citep[e.g., ][]{Henley:al:2003}, and indeed, we do see broad profiles, comparable to the
primary's wind velocity (Figure~\ref{fig:fwhmwav}).  Whether this is
indeed due to the flow in the shock-cone will require spectra at
phases near conjunctions, when such a flow would be transverse to the
line-of-sight, and line widths would be smaller and have different
velocity offsets.

Our plasma model assumes a power-law emission measure distribution.
There is some theoretical justification for this form for stellar
winds \citep{cassinelli:ignace:al:2008,krticka:al:2009,ignace:waldron:al:2012}.  How well a power-law representation applies in colliding winds
is an open question, though it does provide a reasonably good fit to
the spectrum, as seen in Figure~\ref{fig:xrayspec}.  The model has a
parameter specifying a high-temperature cutoff.  The continuum comes
from thermal bremsstrahlung and bound-free processes, and each has an
exponential cutoff at short wavelengths.  Furthermore, the region
below about $10\mang$ has a strong continuum, relatively sparse
emission lines from high-temperature ions, and is least affected by
line-of-sight neutral plasma absorption.  Hence, the high-temperature
cutoff should be a fairly robust indicator of the maximum pre-shock
velocities, under the assumption of strong shocks and collisional
ionization equilibrium emissivities.  Using the maximum temperature of
$50\mk$ (see Table~\ref{tbl:plawdemparams}), a mean molecular weight,
$\mu$ for a highly ionized plasma of $0.7$ appropriate for an
He-enriched plasma (for comparison, abundances of
\citet{Asplund:Grevesse:al:2009} imply $\mu \sim 0.6$), and the
relation for a strong shock, $kT = \frac{3}{16} \mu m_\mathrm{H} v^2$,
in which $T$ is the post-shock temperature, $v$ is the pre-shock
velocity, and $k$ is Boltzmann's constant, we find a maximum relative
shock velocity $v\approx1918\kms$ (with a formal statistical
uncertainty of $66\kms$). \citet{arora:al:2019} estimated that at
$\phi=0.9$ each wind has a velocity of about $1800\kms$ at the
collision, so we might expect anything up to $3600\kms$.  Given that
the emergent spectrum is from a larger region where plasma is cooling,
our estimate is a plausible value.


\subsection{Line Shapes}

Our models, plasma or line-groups, have so far assumed Gaussian
profiles.  There is no {\em a priori} reason for the emission lines being
Gaussian.  The instrument response is highly Gaussian, but these are
resolved features, with a $FWHM$ of several resolution elements.
\citet{Henley:al:2003} have shown how colliding wind regions could
produce a wide variety of profiles, from narrow and single-peaked, to
broad and double-peaked, suggesting a careful review of the observed
line shapes.

The best, relatively isolated and strong feature in the \wrtf spectrum
is that of the H-like \eli{Mg}{12}.  This is an unresolved doublet
with emissivities in a two-to-one ratio: $8.4192\mang$ and
$8.4246\mang$, so the flux-weighted wavelength is $8.4210\mang$.  The
only significant nearby feature, according to the plasma model, is
\eli{Fe}{23} $8.3038\mang$, with a flux at least 15 times smaller.  To
investigate whether lines are asymmetric, we fit the \eli{Mg}{12}
region profiles with a Weibull distribution, whose shape can range
from exponential, to Gaussian-like, to asymmetric with a tail on
either the high or low side of the maximum:
\begin{equation}
  f( \lambda; q, \sigma, \lambda_0) = 
  \frac{q}{\sigma} \,
  \left( \frac{\lambda - \lambda_0} {\sigma} \right)^{q-1} \,
  \exp\left[ -\left( \frac{ \lambda - \lambda_0}{\sigma} \right)^q \right]
\end{equation}\label{eq:weibull}
in which $q>0$ is a shape parameter, $\sigma > 0$ is a the scale, and
$\lambda_0$ is an offset.  We use the mode (maximum) as the line
position:
\begin{equation}
  \lambda_\mathrm{max} =  \lambda_0  +  \sigma \left( \frac{q - 1}{q}\right) ^ \frac{1}{q}
\end{equation}\label{eq:weibullmode}

\begin{figure}
  \centering\leavevmode
  \includegraphics*[width=0.45\columnwidth]{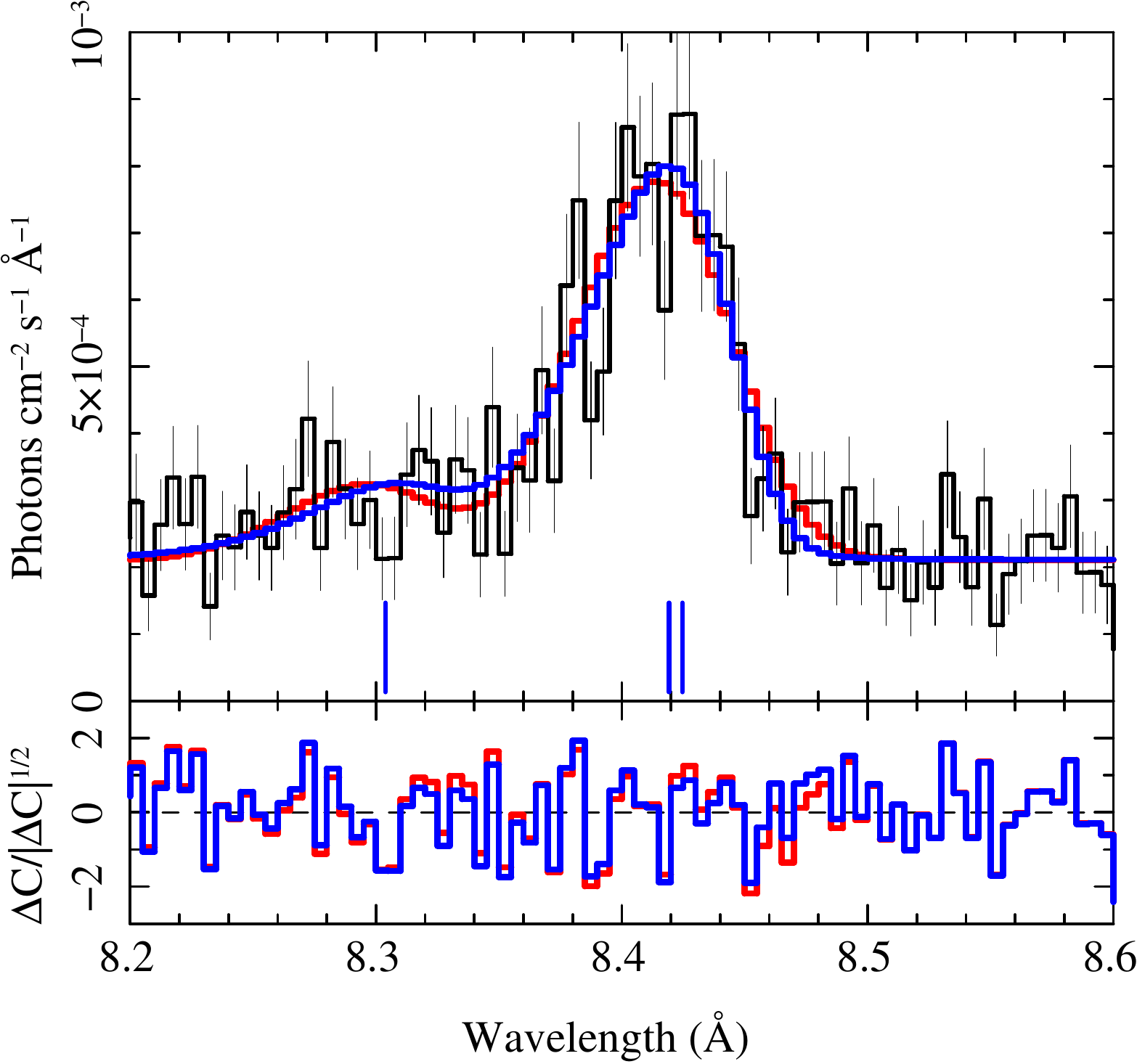}
  \includegraphics*[width=0.45\columnwidth]{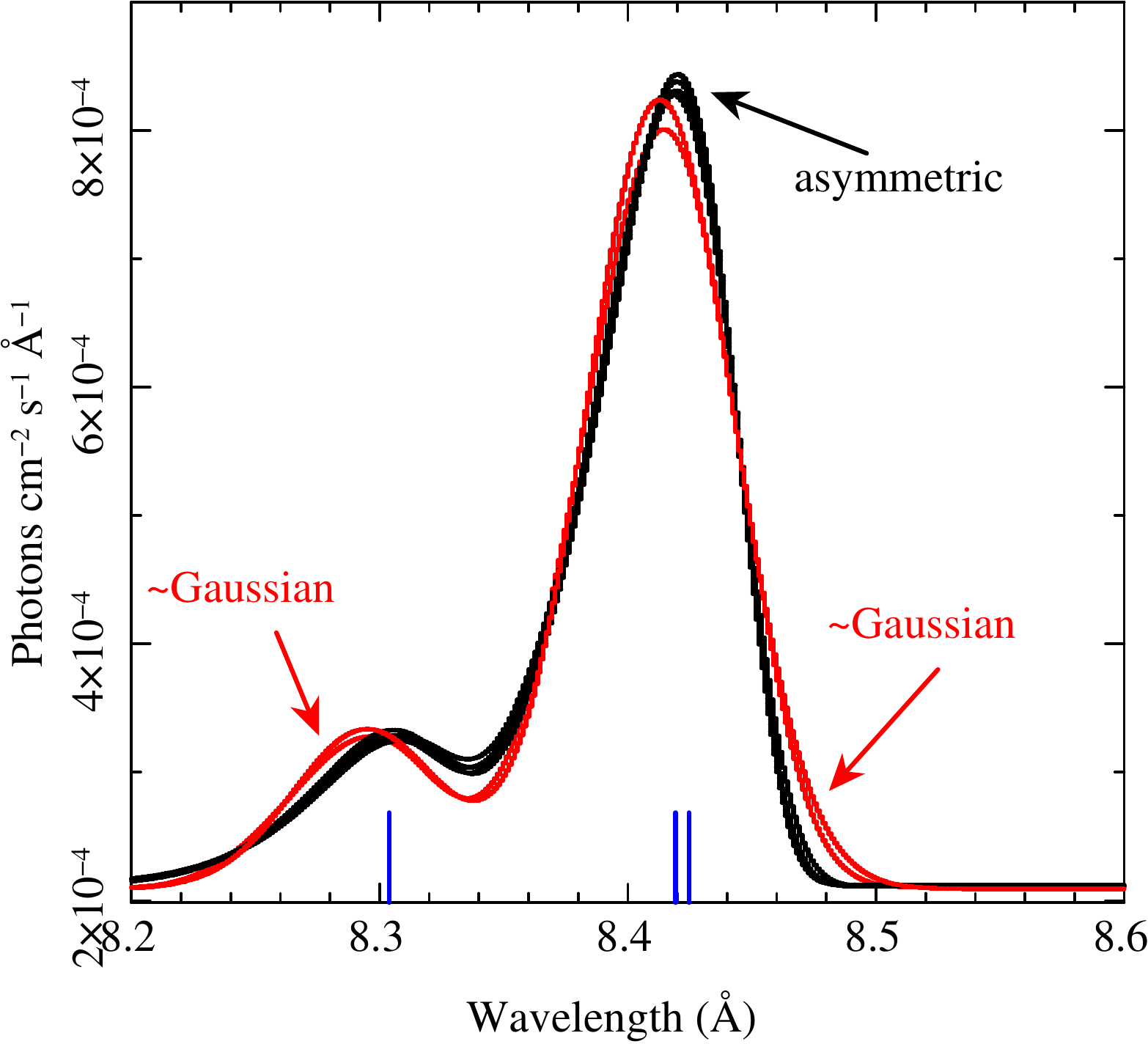}
  \caption{Left:
  The \eli{Mg}{12} region and fit with a Weibull distribution. The
  black histogram shows the data, the colored histograms show two
  models, and the lower panel the respective residuals. The strong feature 
  is the \eli{Mg}{12} doublet at $8.42\mang$; the weak feature at 
  $8.3\mang$ is an \eli{Fe}{23} line.  The red model
  curve is for the $q=4$ (near Gaussian case), and blue is for $q=42$
  (preferred solution).  The blue residuals (lower panel) are
  slightly, but systematically, less than the red.  Differences are
  very subtle, but slightly prefer the asymmetric model.  Right: model profiles 
  (unconvolved by the instrumental profile), all of them
  acceptable, for the $q=4$ (nearly Gaussian) and a Guassian fit (red
  curves, labeled ``$\sim$Gaussian''), and for the slightly preferred
  Weibull fits with larger $q$ (black, labeled ``asymmetric''). These
  serve as a key to the model curves in the left panel.  The long
  vertical tic-marks on the $x$-axes in each plot mark the line rest
  wavelengths.}
  \label{fig:mgfit}
\end{figure}

The model we fit was the sum of two Weibull functions (one for
\eli{Mg}{12} and one for \eli{Fe}{23}) with common $q$ and $\sigma$
parameters, the wavelength offset and the relative strengths of the
lines were constrained to their theoretical differences, plus a
constant continuum, over the $8.0$--$9.0\mang$ region.  Given the best
fit, we then determined confidence intervals via {\tt emcee}, which
also showed that $q$ and $\sigma$ were strongly correlated, with $q>4$
($\sigma>0.1)$, but with values as large as $q=40$ preferred (and
corresponding $\sigma=1.2$).  The best fit has an asymmetric line,
with a slightly more extended wing on the blue side.  However, the
acceptable value of $q=4$ is nearly Gaussian.  The primary difference
is that the peak is closer to the rest wavelength for the Weibull fit
than for the Gaussian, which is probably offset to compensate for the
slight asymmetry.  Statistically, it is challenging to distinguish the
difference, but the notion of some asymmetry agrees with
expectations of more absorption at receding velocities.  We show the
fit and model profiles in Figure~\ref{fig:mgfit} and model parameters
in Table~\ref{tbl:mgparams}.
%
\begin{deluxetable}{crrc}
  \tablecaption{\eli{Mg}{12} Region Weibull Fit Parameters}
  \tablehead{
    \colhead{Parameter}&
    \colhead{Value}&
    \colhead{stddev}&
    \colhead{Units}\\
  }
  \startdata
  $f($\eli{Mg}{12}$)$& 
  $4.60$&
  $0.55$&
  $10^{-5}\,\mathrm{phot\,cm^{-2}\,s^{-1}}$\\
  $\lambda_\mathrm{max}$&
  $8.4198$&
  $0.002$&
  $\mang$\\
  $\Delta v$&
  $-43.7$&
  $79.7$&
  $\kms$\\
  $\sigma$&
  $1.16$&
  $0.40$&
  \\
  $q$&
  $42.5$&
  $13.5$&
  \\
  $f($\eli{Fe}{23}$)$&
  $0.70$&
  $0.19$&
  $10^{-5}\,\mathrm{phot\,cm^{-2}\,s^{-1}}$
  \\
  $f_c$&
  $20.03$&
  $0.55$&
  $10^{-5}\,\mathrm{phot\,cm^{-2}\,s^{-1}\,\mang^{-1}}$
  \\
  \enddata
  \tablecomments{``stddev'' is the one standard
    deviation uncertainty.  The \eli{Fe}{23} line
    shared the parameters $\sigma$ and $q$, and was at a fixed
    wavelength offset from \eli{Mg}{12} of $-0.1181\mang$.}
\end{deluxetable}\label{tbl:mgparams}


An alternative approach has also clearly suggested asymmetry of the line profile.
Following \citet{pollock:2007}, the HETG spectra at maximum light were fitted with an ensemble
of lines with a single
triangular velocity profile above an absorbed Bremsstrahlung continuum. A comparison was made between
symmetric and asymmetric models, the latter parameterized by blue, central and red velocities
showing where the profile reaches zero on either side of a central peak.
The asymmetric model resulted in a \texttt{C-Statistic} improvement
of ${\Delta}C=55.7$ for the single extra degree of freedom over the constrained
symmetric model, clear evidence of a departure from symmetry
with best-fit velocities having the blue intercept of
$-2426\pm103 \kms$, the peak occurring at 
$+61\pm85 \kms$, and a red intercept of
$+1525\pm80 \kms.$

\subsection{He-like Line Ratios}

The He-like atomic structure has metastable levels which produce 
resonance ($r$), intercombination ($i$), and forbidden ($f$) lines in
the soft X-ray band for several abundant elements, making them
excellent diagnostics for temperature, density, and  UV
photoexcitation  \citep{Gabriel:69, Blumenthal:Drake:al:1972}.
These have been used to estimate the radius of formation of X-rays in
OB-star winds \citep{Waldron:Cassinelli:2001, Leutenegger:al:2006,
  Waldron:Cassinelli:2007}.  An intense UV field can depopulate the
forbidden line level, since they are metastable, and the closer the
emitting plasma is to the UV-strong photosphere, the smaller the $f/i$
flux ratio.  High density in plasmas can produce the same effect \citep[c.f., ]{2017ApJ...845...39O}.

Since the X-ray flux of \wrtf comes predominantly from the shock cone,
relatively far from the WR and OB star, we might not expect the $f/i$
ratio for Ne, Mg, and Si (for which we have good resolution) to be
affected, unless there is a significant diffuse
UV field produced in the shocks themselves.  We have used our plasma model as a
baseline for investigating the $f/i$ ratios in \eli{Ne}{9},
\eli{Mg}{12}, and \eli{Si}{13}.  Starting with this model, we
introduce modified emissivities for the He-like lines using data
produced from APEC \citep{foster2012}.\footnote{See
  \url{https://space.mit.edu/cxc/analysis/he_modifier/index.html} for
  data tables, a description, and an implementation for ISIS.}  To let
the model adjust locally to each He-like region investigated, we
allowed the normalization, Doppler shift, line widths, and elemental
abundances to be free, as well as the $f/i$ ratio parameter (we used
density as a simple parameter to scale the ratio).

It is common to refer the ratio of the forbidden line flux to the intercombination line flux as, $R=f/i$.  
The expected value without UV or collisional pumping is usually denoted as $R_0$.  This quantity has a mild temperature
dependence, with about a $10\%$ increase with temperature where the line has significant emissivity.  For $R_0$, we adopted the 
mean values as weighted by emission measure according to our adopted emission measure distribution model (see Section~\ref{sec:onax} and Table~\ref{tbl:plawdemparams}).

For \eli{Mg}{11}, we found that $R/R_0$ $>0.3$ (90\% confidence) or $>0.4$ (68\%
confidence).  \eli{Ne}{9}, at the 90\% confidence level, spans the
entire range; at 68\%, it has $R/R_0>0.25$.  The best fit for \eli{Si}{13}
is consistent with the unexcited limit; the 90\% and 68\% lower limits
to $R/R_0$, respectively, are $0.65$ and $0.76$.  Upper limits for Ne,
Mg, and Si were all $1.0$, or in other words, they are consistent with
no photo-excitation.

\subsection{Variability}\label{sec:var}

The Swift/XRT light curve indicates that the peak emission occurs around
10 days before  periastron. The 0.3-1.5\,keV XRT lightcurves
exhibit a steep drop in the X-rays indicating increased absorption
causing extinction of soft X-rays just after this peak and continues
up to 10 days after periastron (Fig.~\ref{fig:swift}).

The on-axis HETG observations were carried out just around maximum
light as seen in the schematic diagram of the \wrtf~orbit in
Fig.~\ref{fig:orb}. The times of these two observations are also indicated
by green stars in Fig.~\ref{fig:swift}. 

The off-axis HETG spectra, while they are of too low a resolution for
line shape or width determinations, do provide some spectral
information.  Further, the line fluxes and overall energy distribution provide
some diagnostics of the plasma.  We fit these spectra with the same
absorbed powerlaw emission measure model as used for the on-axis
spectra, and also assessed parameter confidence with {\tt emcee};
model parameters are given in Table~\ref{tbl:plawdemparams}, and the
spectra are shown in Figure~\ref{fig:xrayspec}.  By $\phi=0.05$, the
observed (absorbed) flux has dropped by about 40\%, while the maximum
temperature has risen from  $50\mk$ to  $85\mk$.  The
absorbing column has also risen; our sightline passes through more
of the WR-star's wind.  The temperature increase is
unexpected, since the stellar separations are about the same at the two phases, hence the wind-wind collision should be at the same relative velocity.  At
$\phi=0.13$, the absorption has dropped substantially as expected; 
the observed flux is about the same as at phase $0.05$.  Here, we might expect the somewhat higher temperature as seen, given that the winds are expected
to collide at a somewhat higher relative velocity.  At the higher
maximum plasma temperature, we estimate a shock relative velocity of
$\sim2200\kms$.

The unabsorbed fluxes decrease steadily over the phase range from $0.95$ to $0.13$.  We expect flux to generally scale inversely with the
separation of the stars.  Since it does not strictly do this, it may indicate other than adiabatic processes, as suggested by \citet{arora:al:2019}, who estimated that radiative processes become important near periastron.

\begin{figure}[ht!]
  \centering\leavevmode
  \includegraphics*[width=0.48\columnwidth, angle=0]{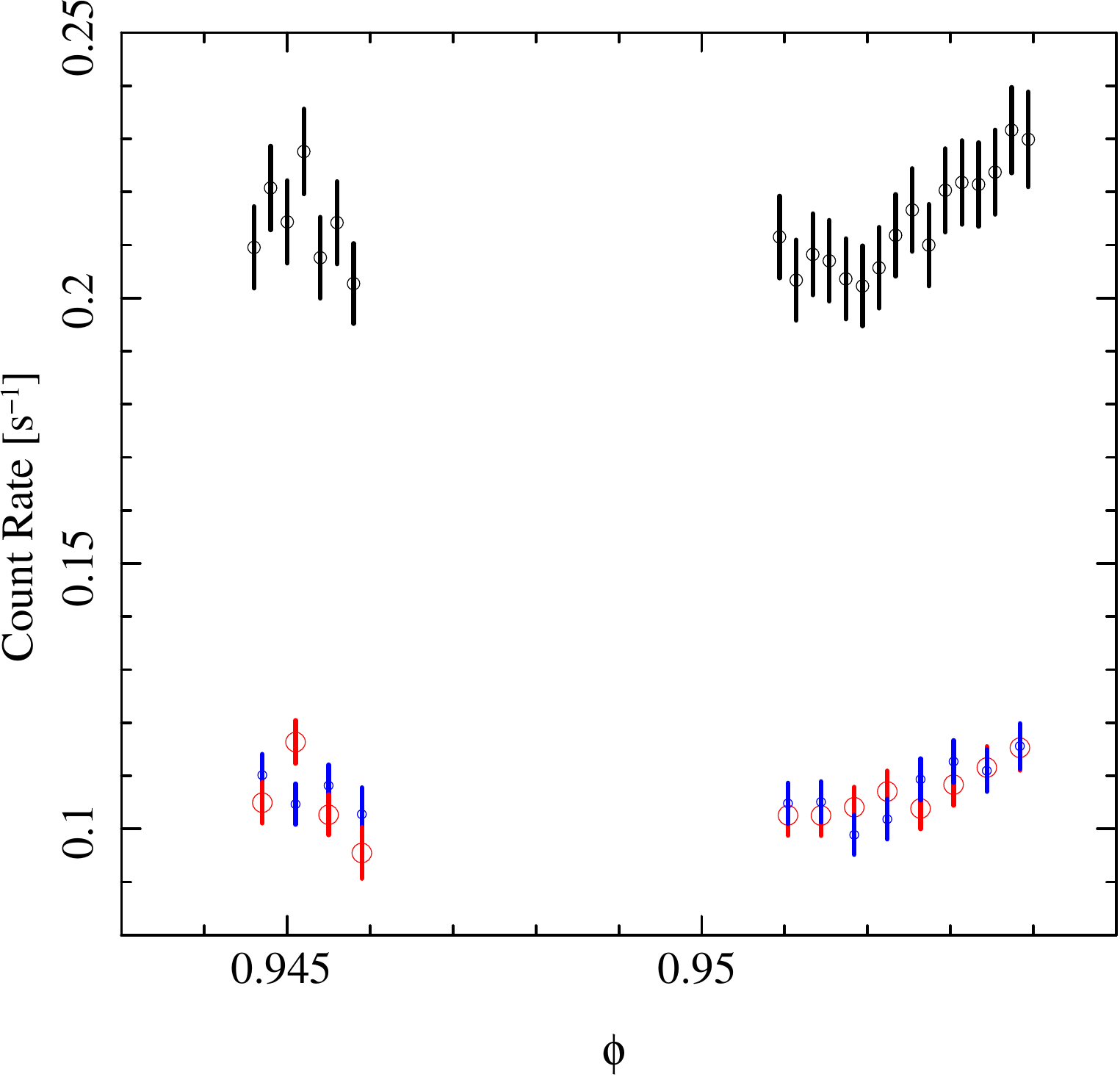}
  \includegraphics*[width=0.48\columnwidth, angle=0]{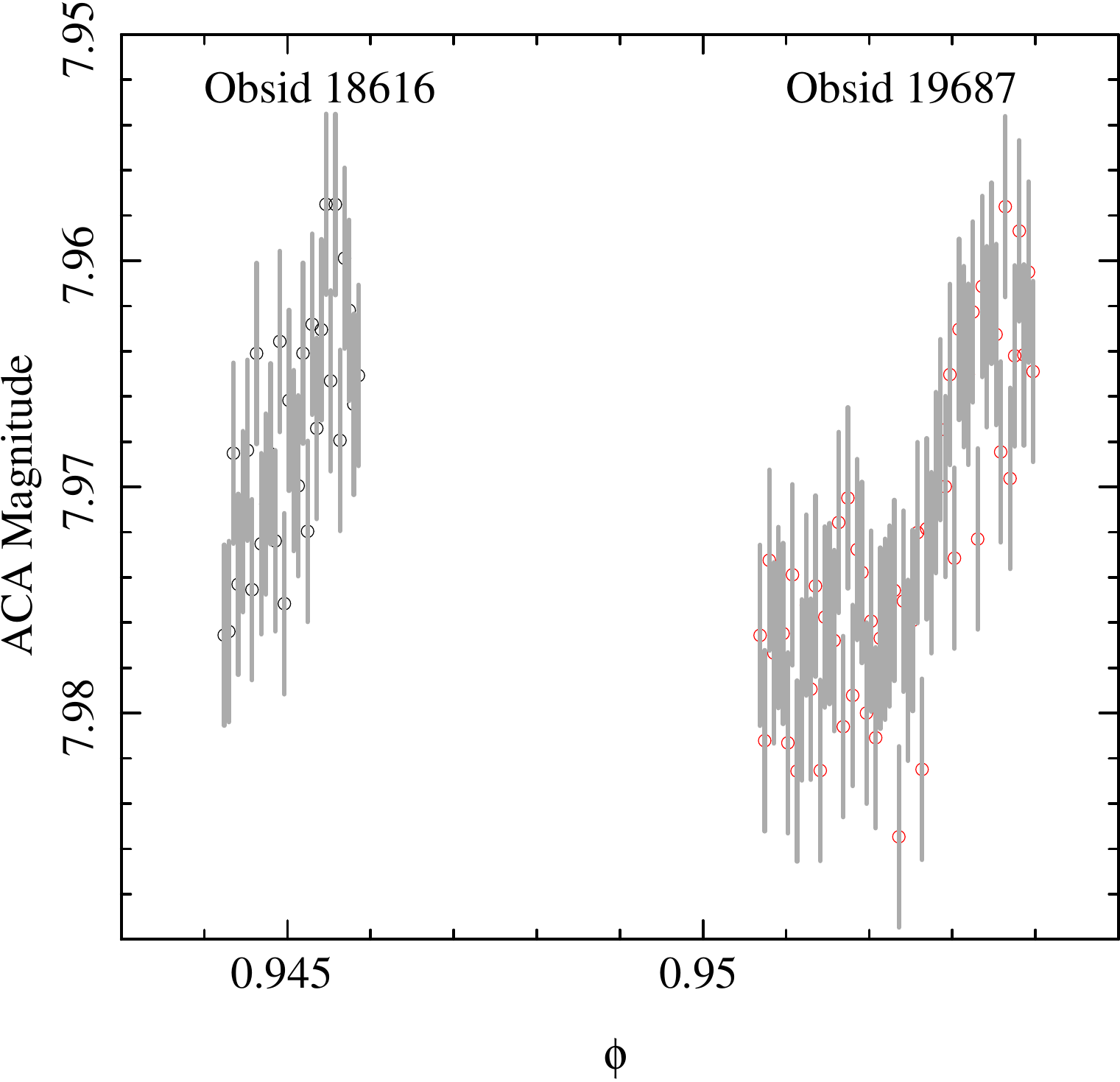}
  \caption{Left: We show the count-rate light curve for the on-axis pointings. The top
    shows the $1.7$ -- $17.0\mang$ band for HEG and MEG first orders.
    Below the curves are shown for two bands, subdivided at $7.25\mang$;
    small (blue) circles are the short-wavelength range, and larger
    circles (red) the long-wavelength range. There is no appreciable
    change in the hardness ratio for these two bands.  Error bars are one
    standard deviation, and time bins were about 1 hour.  The $x$-axis range spans about $2.5\,$days at about $10\,$days before periastron. Right: 
    The ACA magnitudes and estimated errors (see text) of \wrtf for ObsIDs\dataset[18616]{ https://heasarc.gsfc.nasa.gov/FTP/chandra/data/science/ao17/cat2//18616/} and \dataset[19687]{https://heasarc.gsfc.nasa.gov/FTP/chandra/data/science/ao17/cat2//19687/}.  \wrtf was monitored with the optical ACA during these observations.  The data are binned at 1\,ks.  The phase of the observations is shown on the X axis.
  } 
  \label{fig:xrlctwo}
\end{figure}

Within the pointed observation, there was significant short-term
variability.  In Figure~\ref{fig:xrlctwo} (left),  we show the light curve of
dispersed X-ray photons.  There is significant variability, but it is not a
constant rise with phase as the system approaches periastron.  This
could represent non-uniform density structure in the wind.

 During the two on-axis observations, obsids 18616 and 19687, \wrtf was monitored optically with the Aspect Camera onboard Chandra.  The Aspect Camera is mounted parallel to the X-ray telescope boresight and read continuously during the observation approximately every 2.05 s.  The Aspect Camera CCD has a spectral response from 4000-10000 \AA\, so it is a redder wavelength range than Johnson UBV photometry.  Flux calibration is based on a zero magnitude star of G0 V spectral type.  Conversion between V and B magnitudes and Aspect Camera magnitudes is available in \citet{nichols2010}.  It has been shown that the Aspect Camera photometry is internally consistent to 1-2\%.

During the two on-axis observations, obsids 18616 and 19687, WR25 was monitored optically with the Aspect Camera onboard Chandra.  The Aspect Camera is mounted parallel to the X-ray telescope boresight and read during the observation approximately every 2.05 s.  The Aspect Camera CCD has a spectral response from 4000-10000 \AA\, so it is a redder wavelength range than Johnson UBV photometry.  Flux calibration is based on a zero magnitude star of G0 V spectral type.  Conversion between V and B magnitudes and Aspect Camera magnitudes is available in Nichols (2010).  It has been shown that the Aspect Camera photometry is internally consistent to 1-2\%.

Figure \ref{fig:xrlctwo} (right) shows the Aspect Camera (ACA) data for ObsIDs\dataset[18616]{ https://heasarc.gsfc.nasa.gov/FTP/chandra/data/science/ao17/cat2//18616/} and \dataset[19687]{https://heasarc.gsfc.nasa.gov/FTP/chandra/data/science/ao17/cat2//19687/} plotted with \wrtf phase and the ACA magnitude.  It is surprising to find both light curves show a similar rise in brightness (lower magnitude).  We verified that no high background event took place during the acquisition of these data, and all guide stars were well tracked during the entire observation (E. Martin, private communication).  To estimate the error in ACA magnitude,  we used the mean of the standard deviations of the ACA magnitudes for two of the guide stars used in this observation, deriving a value of 0.004 ACA magnitude.  The two guide stars selected were clearly contant in ACA magnitude, so the errors we show for the ACA magnitude of \wrtf are assumed to be the errors associated with a constant source.  The ACA data are broadband, with no spectral distribution information, so we cannot determine the cause of the change in magnitude during each observation.  One possibility is a change in reddening, perhaps due to clumps of dust or density-enhanced gas clouds in the line of sight.  \wrtf in fact lies in the line of sight to an enhanced thermal dust feature in Planck data.  \wrtf has no intrinsic polarization \citep{fullard2020} but a large IPS angle (deviation of polarization angle with wavelength).  \citet{drissen1992} suggested that this deviation could be due to regions within the Carina Nebula that experience grain processing due to shock waves and the large IPS angle determined for  \wrtf \citep{fullard2020} supports this idea.  However, intrinsic dust in the  \wrtf system could also produce a dust absorption signature.  Optical and IR monitoring would be required to study the time-dependent absorption in this line-of-sight.

\section{Conclusions}\label{sec:conc}

We have presented the first high-resolution X-ray spectra of the
colliding wind binary \wrtf, along with the context of long-term
broad-band X-ray monitoring with Swift.  The pointed \hetgs
observations were near periastron, when the X-ray brightness is
maximal.  The colliding wind shock region dominates the emission,
being about an order of magnitude above that from similar spectral
type stars which lack wind-wind collisions.

Our primary results from fitting the high-resolution lines and the
energy distribution, under the assumption that the emitting plasma is
in collisional ionization equilibrium,  are:
\begin{enumerate}
\item The pre-shock velocity is about $v\approx1918\kms$;
\item Emission lines have a blue-shifted centroid which increases in
  magnitude with increasing wavelength, from about $-100\kms$ to about
  $-600\kms$.  This is understood as likely due to the increased
  opacity of neutral matter which increasingly absorbs emission from
  receding plasma.
\item Emission lines are resolved and broad, with a $FWHM$ of about
  $2400\kms$, with no clear trend with wavelength.  This is consistent
  with the velocities of the colliding stellar winds seen nearly
  along the line-of-sight in a shock-cone.   The constancy of the width with
  wavelength is somewhat at odds with the previous item, since
  increasing opacity with wavelength would hide receding material and
  narrow the lines.  There might be significant turbulence in the
  shock-cone which masks bulk flow effects.
\item The emission lines seem very nearly Gaussian, though formally, a
  slightly asymmetric profile is preferred.  The line shape depends
  critically on velocity fields and opacity.  Hence, this should
  provide an important constraint for any detailed hydrodynamical
  models. 
\item Within the \hetgs observation, the source X-ray flux was
  significantly variable, at a level of about $15\%$.  While phase
  coverage was small, the changes seemed gray and non-secular with
  phase.  They may represent clumpiness in the wind and not, for
  instance, changing absorption or temperature.
\end{enumerate}

The off-axis \hetgs spectra, while higher resolution than
imaging-mode, are blurred enough that line centroid and profile
information are not available.  The behavior does not show simple
dependence on stellar separation, or phase, and may indicate that
there non-adiabatic processes are significant near periastron.
High resolution spectra are required at these phases in order to
determine the dynamics, and to provide crucial constraints on
wind-shock models.  For instance, at $\phi\sim0.04$ (conjunction), we
expect the shock cone flow to be primarily away from us we would
expect redshifted lines.  The line width should also be smaller than
near periastron since we are not viewing the cone transversely, but
face on.   At the other conjunction (O-star in front; $\phi\sim0.8$,
$-40\,$days), the shock cone would be concave toward the viewer, and
we would expect blue-shifted lines.

\wrtf is perhaps the best system for studying colliding-wind physics.
It is bright, has a period amenable to phase-resolved studies in a
relatively short time span, and most importantly, it appears to stay
relatively transparent to the O4-star even when looking through the WN
6 star's wind.  Other systems, such as WR 140 ($P\sim8\,\mathrm{yrs}$)
or $\eta\,$Car ($P\sim5\,\mathrm{yrs}$) are quite opaque during
conjunction and do not allow the shock-cone to be seen at soft X-ray
energies.  Blue-shifted lines have regularly been observed in
colliding-wind binaries, notably in WR~140, but red-shifts only in
$\theta\,$Muscae, a complex multiple system where orbital dynamics
seem to play no role.  In WR~140, absorption by the WC star wind at
conjunction is far too strong for measurements to be feasible. The
very massive WN systems Mk~34 and WR~21a both show deep extended
minima at their relevant conjunction phases.  While these are all
interesting and relevant systems, \wrtf remains the best for studies
of shock-cone dynamics about the whole orbit, and hence an excellent
subject for study of stellar winds.

This study can only be considered as the initial investigation into
the dynamics, since more phase coverage is needed.  But we have shown
that such a study is feasible and would likely be fruitful with the
\chan/\hetgs.

\begin{acknowledgements}
Support for this work was in provided by NASA through the Smithsonian
Astrophysical Observatory (SAO) contract SV3-73016 to MIT for Support
of the Chandra X-Ray Center (CXC) and Science Instruments, and by
Chandra Award Number AR8-19001 (A, B, and C) issued by the CXC.  
The CXC is operated by the Smithsonian Astrophysical Observatory 
for and on behalf of NASA under contract NAS8-03060. JSN also acknowledges support of Chandra contract NAS8-03060. The unique facilities offered by the Swift Target of Opportunity program and the UKSSDC pipeline
were crucial and we thank the Swift Project Scientist and colleagues.

\end{acknowledgements}

\appendix

The detailed plots for the Gaussian fits for two on-axis observations (Obsids 18616 and 19687) with Chandra/HETGS are shown here. 
\renewcommand{\thefigure}{A\arabic{figure}}
\setcounter{figure}{0}

\begin{figure}[htb]
  \centering\leavevmode
  \includegraphics[width=1.0\columnwidth,angle=0]{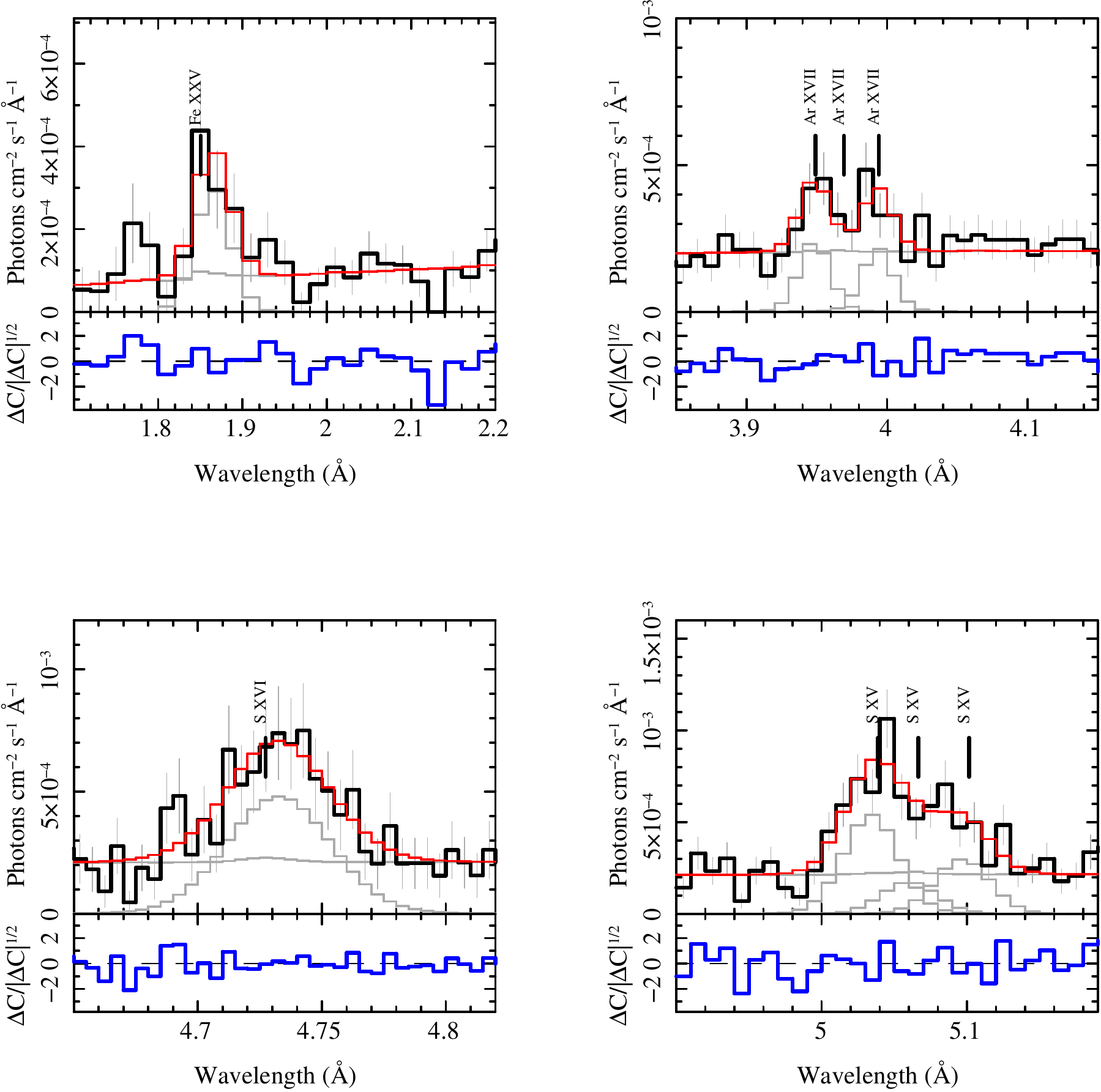}
  \caption{Gaussian fits to several emission line regions for the on-axis observations. In each panel, the black histogram shows the flux spectrum for the combined HETG first orders, the red is the model, the grey are the individual components of the model and the blue are the residuals.}
  \label{fig:lineprof1}
\end{figure} 

\begin{figure}
  \centering\leavevmode
  \includegraphics[width=1.0\columnwidth,angle=0]{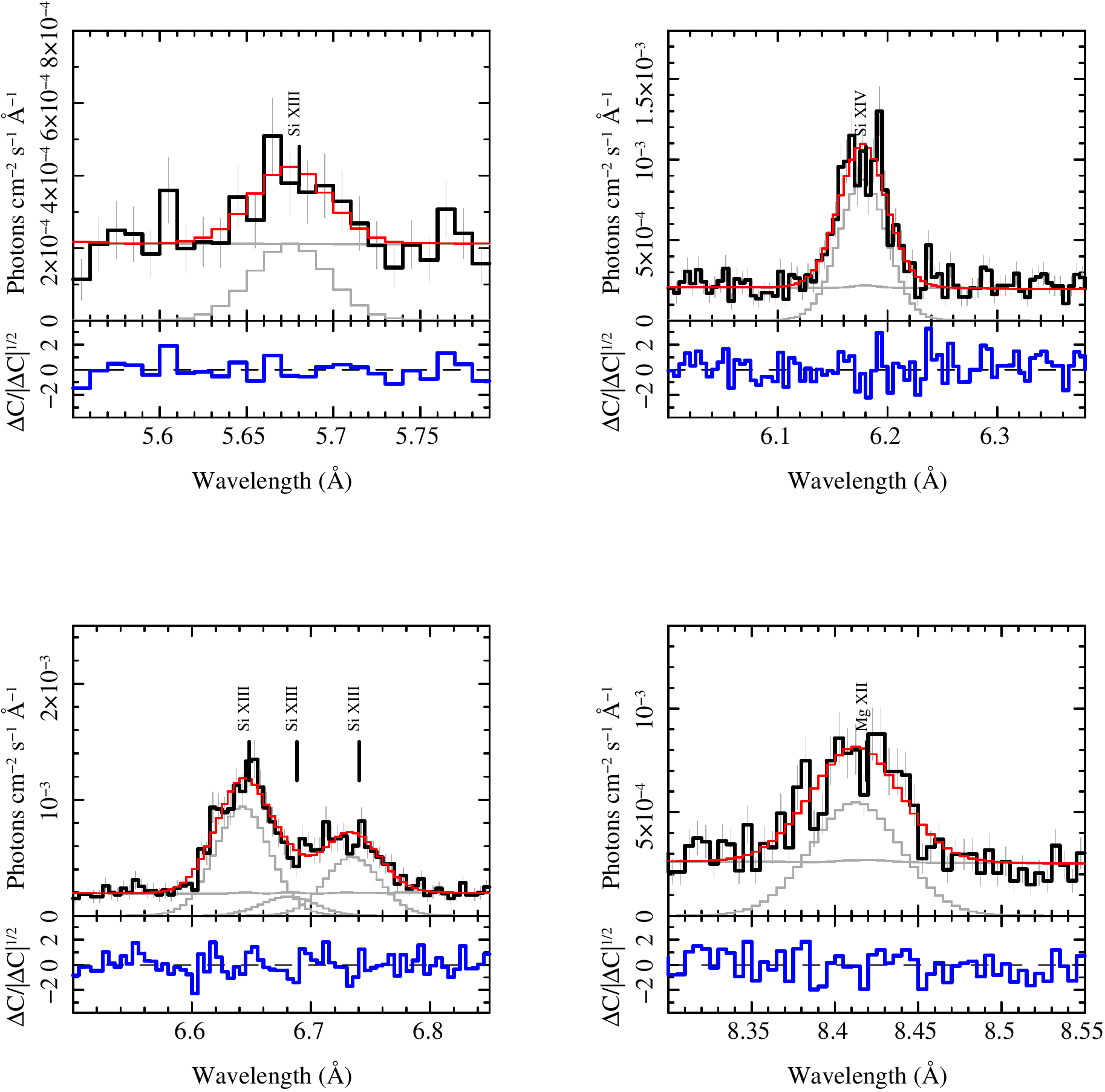}
  \caption{Gaussian fits to several emission line regions for the on-axis observations. In each panel, the black histogram shows the flux spectrum for the combined HETG first orders, the red is the model, the grey are the individual components of the model and the blue are the residuals.}
  \label{fig:lineprof2}
\end{figure} 

\begin{figure}
  \centering\leavevmode
   \includegraphics[width=1.0\columnwidth,angle=0]{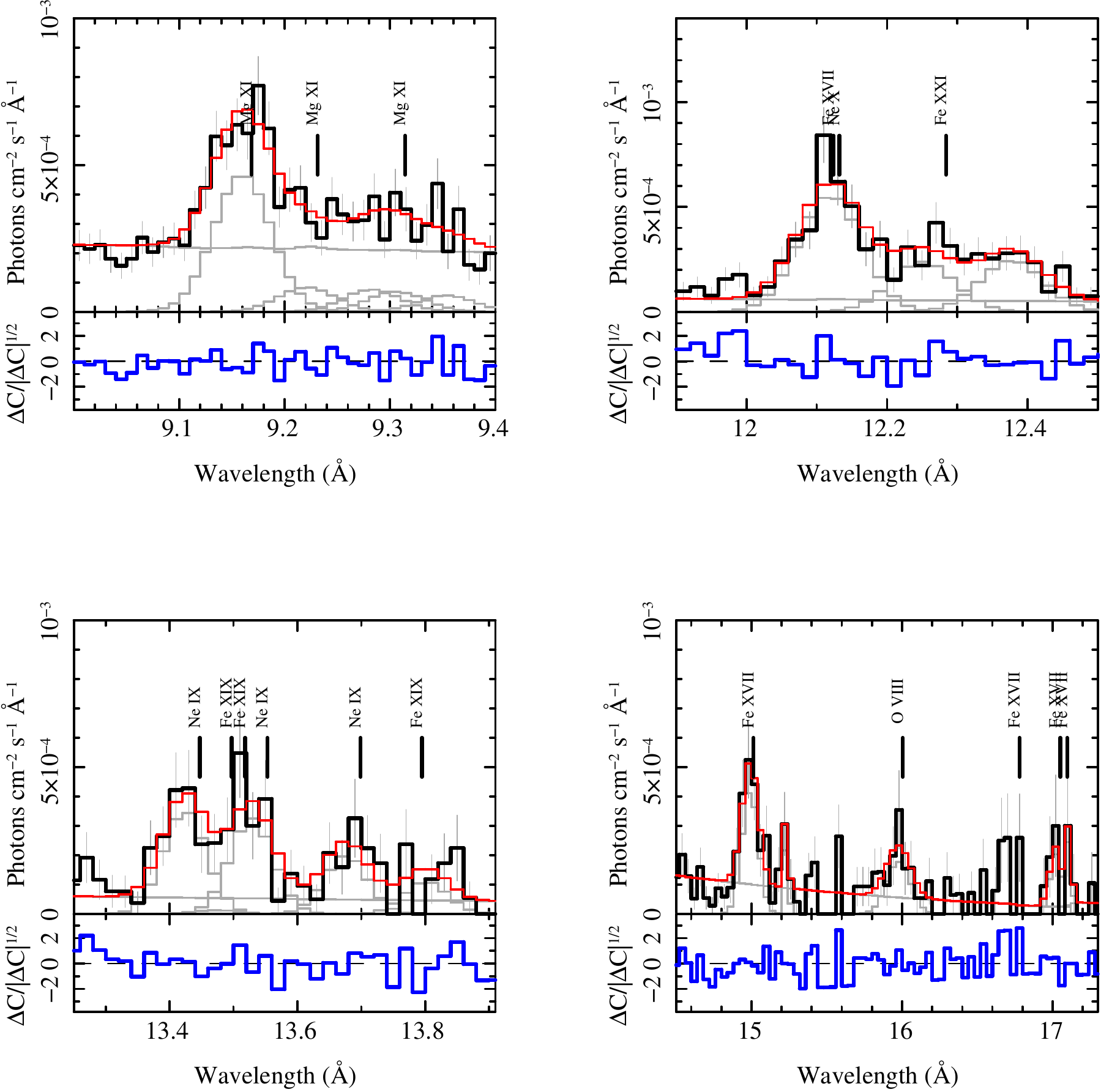}
  \caption{Gaussian fits to several emission line regions for the on-axis observations. In each panel, the black histogram shows the flux spectrum for the combined HETG first orders, the red is the model, the grey are the individual components of the model and the blue are the residuals.}
  \label{fig:lineprof3}
\end{figure} 


%

%

\clearpage

\bibliographystyle{aasjournal} 
\bibliography{wr25}     

\begin{thebibliography}{}
\expandafter\ifx\csname natexlab\endcsname\relax\def\natexlab#1{#1}\fi
\providecommand{\url}[1]{\href{#1}{#1}}
\providecommand{\dodoi}[1]{doi:~\href{http://doi.org/#1}{\nolinkurl{#1}}}
\providecommand{\doeprint}[1]{\href{http://ascl.net/#1}{\nolinkurl{http://ascl.net/#1}}}
\providecommand{\doarXiv}[1]{\href{https://arxiv.org/abs/#1}{\nolinkurl{https://arxiv.org/abs/#1}}}

\bibitem[{{Arnaud}(1996)}]{Arnaud:1996}
{Arnaud}, K.~A. 1996, in Astronomical Society of the Pacific Conference Series,
  Vol. 101, Astronomical Data Analysis Software and Systems V, ed. G.~H.
  {Jacoby} \& J.~{Barnes}, 17--+

\bibitem[{{Arora} {et~al.}(2019){Arora}, {Pandey}, \& {De
  Becker}}]{arora:al:2019}
{Arora}, B., {Pandey}, J.~C., \& {De Becker}, M. 2019, \mnras, 487, 2624,
  \dodoi{10.1093/mnras/stz1447}

\bibitem[{{Asplund} {et~al.}(2009){Asplund}, {Grevesse}, {Sauval}, \&
  {Scott}}]{Asplund:Grevesse:al:2009}
{Asplund}, M., {Grevesse}, N., {Sauval}, A.~J., \& {Scott}, P. 2009, \araa, 47,
  481, \dodoi{10.1146/annurev.astro.46.060407.145222}

\bibitem[{{Blumenthal} {et~al.}(1972){Blumenthal}, {Drake}, \&
  {Tucker}}]{Blumenthal:Drake:al:1972}
{Blumenthal}, G.~R., {Drake}, G.~W.~F., \& {Tucker}, W.~H. 1972, \apj, 172,
  205, \dodoi{10.1086/151340}

\bibitem[{{Cant{\'o}} {et~al.}(1996){Cant{\'o}}, {Raga}, \&
  {Wilkin}}]{1996ApJ...469..729C}
{Cant{\'o}}, J., {Raga}, A.~C., \& {Wilkin}, F.~P. 1996, \apj, 469, 729,
  \dodoi{10.1086/177820}

\bibitem[{{Cassinelli} {et~al.}(2008){Cassinelli}, {Ignace}, {Waldron}, {Cho},
  {Murphy}, \& {Lazarian}}]{cassinelli:ignace:al:2008}
{Cassinelli}, J.~P., {Ignace}, R., {Waldron}, W.~L., {et~al.} 2008, \apj, 683,
  1052, \dodoi{10.1086/589760}

\bibitem[{{Castor} {et~al.}(1975){Castor}, {Abbott}, \& {Klein}}]{CAK}
{Castor}, J.~I., {Abbott}, D.~C., \& {Klein}, R.~I. 1975, \apj, 195, 157

\bibitem[{{Corcoran} {et~al.}(2017){Corcoran}, {Liburd}, {Morris}, {Russell},
  {Hamaguchi}, {Gull}, {Madura}, {Teodoro}, {Moffat}, {Richardson}, {Hillier},
  {Damineli}, \& {Groh}}]{corcoran:al:2017}
{Corcoran}, M.~F., {Liburd}, J., {Morris}, D., {et~al.} 2017, \apj, 838, 45,
  \dodoi{10.3847/1538-4357/aa6347}

\bibitem[{{Crowther}(2007)}]{crowther:2007}
{Crowther}, P.~A. 2007, \araa, 45, 177,
  \dodoi{10.1146/annurev.astro.45.051806.110615}

\bibitem[{{Crowther} \& {Walborn}(2011)}]{crowther:walborn:2011}
{Crowther}, P.~A., \& {Walborn}, N.~R. 2011, \mnras, 416, 1311,
  \dodoi{10.1111/j.1365-2966.2011.19129.x}

\bibitem[{{Driessen} {et~al.}(2019){Driessen}, {Sundqvist}, \&
  {Kee}}]{2019A&A...631A.172D}
{Driessen}, F.~A., {Sundqvist}, J.~O., \& {Kee}, N.~D. 2019, \aap, 631, A172,
  \dodoi{10.1051/0004-6361/201936331}

\bibitem[{{Drissen} {et~al.}(1992){Drissen}, {Robert}, \&
  {Moffat}}]{drissen1992}
{Drissen}, L., {Robert}, C., \& {Moffat}, A. F.~J. 1992, \apj, 386, 288,
  \dodoi{10.1086/171014}

\bibitem[{{Feldmeier} {et~al.}(1997){Feldmeier}, {Puls}, \&
  {Pauldrach}}]{Feldmeier:Puls:Pauldrach:1997}
{Feldmeier}, A., {Puls}, J., \& {Pauldrach}, A.~W.~A. 1997, \aap, 322, 878

\bibitem[{{Foreman-Mackey} {et~al.}(2013){Foreman-Mackey}, {Hogg}, {Lang}, \&
  {Goodman}}]{emcee:2013}
{Foreman-Mackey}, D., {Hogg}, D.~W., {Lang}, D., \& {Goodman}, J. 2013, \pasp,
  125, 306, \dodoi{10.1086/670067}

\bibitem[{{Foster} {et~al.}(2012){Foster}, {Ji}, {Smith}, \&
  {Brickhouse}}]{foster2012}
{Foster}, A.~R., {Ji}, L., {Smith}, R.~K., \& {Brickhouse}, N.~S. 2012, \apj,
  756, 128, \dodoi{10.1088/0004-637X/756/2/128}

\bibitem[{{Fruscione} {et~al.}(2006){Fruscione}, {McDowell}, {Allen},
  {Brickhouse}, {Burke}, {Davis}, {Durham}, {Elvis}, {Galle}, {Harris},
  {Huenemoerder}, {Houck}, {Ishibashi}, {Karovska}, {Nicastro}, {Noble},
  {Nowak}, {Primini}, {Siemiginowska}, {Smith}, \& {Wise}}]{CIAO:2006}
{Fruscione}, A., {McDowell}, J.~C., {Allen}, G.~E., {et~al.} 2006, in Presented
  at the Society of Photo-Optical Instrumentation Engineers (SPIE) Conference,
  Vol. 6270, SPIE Conference Series, \dodoi{10.1117/12.671760}

\bibitem[{{Fullard} {et~al.}(2020){Fullard}, {St-Louis}, {Moffat}, {Piirola},
  {Manset}, \& {Hoffman}}]{fullard2020}
{Fullard}, A.~G., {St-Louis}, N., {Moffat}, A. F.~J., {et~al.} 2020, \aj, 159,
  214, \dodoi{10.3847/1538-3881/ab8293}

\bibitem[{{Gabriel} \& {Jordan}(1969)}]{Gabriel:69}
{Gabriel}, A.~H., \& {Jordan}, C. 1969, \mnras, 145, 241

\bibitem[{{Gamen} {et~al.}(2006){Gamen}, {Gosset}, {Morrell}, {Niemela},
  {Sana}, {Naz{\'e}}, {Rauw}, {Barb{\'a}}, \& {Solivella}}]{gamen:al:2006}
{Gamen}, R., {Gosset}, E., {Morrell}, N., {et~al.} 2006, \aap, 460, 777,
  \dodoi{10.1051/0004-6361:20065618}

\bibitem[{{Gamen} {et~al.}(2008){Gamen}, {Gosset}, {Morrell}, {Niemela},
  {Sana}, {Naz{\'e}}, {Rauw}, {Barb{\'a}}, \& {Solivella}}]{gamen2008}
{Gamen}, R., {Gosset}, E., {Morrell}, N.~I., {et~al.} 2008, in Revista Mexicana
  de Astronomia y Astrofisica Conference Series, Vol.~33, Revista Mexicana de
  Astronomia y Astrofisica Conference Series, 91--93

\bibitem[{{Gayley}(2009)}]{2009ApJ...703...89G}
{Gayley}, K.~G. 2009, \apj, 703, 89, \dodoi{10.1088/0004-637X/703/1/89}

\bibitem[{{Gayley} {et~al.}(1997){Gayley}, {Owocki}, \&
  {Cranmer}}]{gayley:al:1997}
{Gayley}, K.~G., {Owocki}, S.~P., \& {Cranmer}, S.~R. 1997, \apj, 475, 786,
  \dodoi{10.1086/303573}

\bibitem[{{Grant} {et~al.}(2020){Grant}, {Blundell}, \&
  {Matthews}}]{grant:blundell:al:2020}
{Grant}, D., {Blundell}, K., \& {Matthews}, J. 2020, \mnras, 494, 17,
  \dodoi{10.1093/mnras/staa669}

\bibitem[{{Hamaguchi} {et~al.}(2016){Hamaguchi}, {Corcoran}, {Gull},
  {Takahashi}, {Grefenstette}, {Yuasa}, {Stuhlinger}, {Russell}, {Moffat},
  {Sharma}, {Madura}, {Richardson}, {Groh}, {Pittard}, \&
  {Owocki}}]{hamaguchi:al:2016}
{Hamaguchi}, K., {Corcoran}, M.~F., {Gull}, T.~R., {et~al.} 2016, \apj, 817,
  23, \dodoi{10.3847/0004-637X/817/1/23}

\bibitem[{{Hamann} {et~al.}(2019){Hamann}, {Gr{\"a}fener}, {Liermann},
  {Hainich}, {Sander}, {Shenar}, {Ramachand ran}, {Todt}, \&
  {Oskinova}}]{hamann:grafener:al:2019}
{Hamann}, W.~R., {Gr{\"a}fener}, G., {Liermann}, A., {et~al.} 2019, \aap, 625,
  A57, \dodoi{10.1051/0004-6361/201834850}

\bibitem[{{Henley} {et~al.}(2003){Henley}, {Stevens}, \&
  {Pittard}}]{Henley:al:2003}
{Henley}, D.~B., {Stevens}, I.~R., \& {Pittard}, J.~M. 2003, \mnras, 346, 773,
  \dodoi{10.1111/j.1365-2966.2003.07121.x}

\bibitem[{{Houck} \& {Denicola}(2000)}]{houck2000}
{Houck}, J.~C., \& {Denicola}, L.~A. 2000, in Astronomical Society of the
  Pacific Conference Series, Vol. 216, Astronomical Data Analysis Software and
  Systems IX, ed. N.~{Manset}, C.~{Veillet}, \& D.~{Crabtree}, 591

\bibitem[{{Howarth} \& {van Leeuwen}(2019)}]{howarth:vanleeuwen:2019}
{Howarth}, I.~D., \& {van Leeuwen}, F. 2019, \mnras, 484, 5350,
  \dodoi{10.1093/mnras/stz291}

\bibitem[{Huenemoerder {et~al.}(2011)Huenemoerder, Mitschang, Dewey, Nowak,
  Schulz, Nichols, Davis, Houck, Marshall, Noble, Morgan, \&
  Canizares}]{Huenemoerder2011}
Huenemoerder, D.~P., Mitschang, A., Dewey, D., {et~al.} 2011, The Astronomical
  Journal, 141, 129, \dodoi{10.1088/0004-6256/141/4/129}

\bibitem[{{Huenemoerder} {et~al.}(2020){Huenemoerder}, {Ignace}, {Miller},
  {Gayley}, {Hamann}, {Lauer}, {Moffat}, {Naz{\'e}}, {Nichols}, {Oskinova},
  {Richardson}, \& {Waldron}}]{Huenemoerder:al:2020}
{Huenemoerder}, D.~P., {Ignace}, R., {Miller}, N.~A., {et~al.} 2020, \apj, 893,
  52, \dodoi{10.3847/1538-4357/ab8005}

\bibitem[{{Ignace}(2001)}]{ignace:2001}
{Ignace}, R. 2001, \apjl, 549, L119, \dodoi{10.1086/319141}

\bibitem[{{Ignace}(2016)}]{2016AdSpR..58..694I}
---. 2016, Advances in Space Research, 58, 694,
  \dodoi{10.1016/j.asr.2015.12.044}

\bibitem[{{Ignace} {et~al.}(2000){Ignace}, {Oskinova}, \&
  {Foullon}}]{2000MNRAS.318..214I}
{Ignace}, R., {Oskinova}, L.~M., \& {Foullon}, C. 2000, \mnras, 318, 214,
  \dodoi{10.1046/j.1365-8711.2000.03744.x}

\bibitem[{{Ignace} {et~al.}(2012){Ignace}, {Waldron}, {Cassinelli}, \&
  {Burke}}]{ignace:waldron:al:2012}
{Ignace}, R., {Waldron}, W.~L., {Cassinelli}, J.~P., \& {Burke}, A.~E. 2012,
  \apj, 750, 40, \dodoi{10.1088/0004-637X/750/1/40}

\bibitem[{{Kahn} {et~al.}(2001){Kahn}, {Leutenegger}, {Cottam}, {Rauw},
  {Vreux}, {den Boggende}, {Mewe}, \& {G{\"u}del}}]{Kahn:Leutenegger:al:2001}
{Kahn}, S.~M., {Leutenegger}, M.~A., {Cottam}, J., {et~al.} 2001, \aap, 365,
  L312, \dodoi{10.1051/0004-6361:20000093}

\bibitem[{{Krti{\v{c}}ka} {et~al.}(2009){Krti{\v{c}}ka}, {Feldmeier},
  {Oskinova}, {Kub{\'a}t}, \& {Hamann}}]{krticka:al:2009}
{Krti{\v{c}}ka}, J., {Feldmeier}, A., {Oskinova}, L.~M., {Kub{\'a}t}, J., \&
  {Hamann}, W.~R. 2009, \aap, 508, 841, \dodoi{10.1051/0004-6361/200912642}

\bibitem[{{Langer}(2012)}]{langer:2012}
{Langer}, N. 2012, \araa, 50, 107, \dodoi{10.1146/annurev-astro-081811-125534}

\bibitem[{{Leutenegger} {et~al.}(2006){Leutenegger}, {Paerels}, {Kahn}, \&
  {Cohen}}]{Leutenegger:al:2006}
{Leutenegger}, M.~A., {Paerels}, F.~B.~S., {Kahn}, S.~M., \& {Cohen}, D.~H.
  2006, \apj, 650, 1096, \dodoi{10.1086/507147}

\bibitem[{{Lucy} \& {White}(1980)}]{Lucy:White:1980}
{Lucy}, L.~B., \& {White}, R.~L. 1980, \apj, 241, 300

\bibitem[{{Luo} {et~al.}(1990){Luo}, {McCray}, \& {Mac
  Low}}]{1990ApJ...362..267L}
{Luo}, D., {McCray}, R., \& {Mac Low}, M.-M. 1990, \apj, 362, 267,
  \dodoi{10.1086/169263}

\bibitem[{{Muijres} {et~al.}(2011){Muijres}, {de Koter}, {Vink},
  {Krti{\v{c}}ka}, {Kub{\'a}t}, \& {Langer}}]{2011A&A...526A..32M}
{Muijres}, L.~E., {de Koter}, A., {Vink}, J.~S., {et~al.} 2011, \aap, 526, A32,
  \dodoi{10.1051/0004-6361/201014290}

\bibitem[{{Nichols} {et~al.}(2010){Nichols}, {Henden}, {Huenemoerder}, {Lauer},
  {Martin}, {Morgan}, \& {Sundheim}}]{nichols2010}
{Nichols}, J.~S., {Henden}, A.~A., {Huenemoerder}, D.~P., {et~al.} 2010, \apjs,
  188, 473, \dodoi{10.1088/0067-0049/188/2/473}

\bibitem[{{Oskinova} {et~al.}(2017){Oskinova}, {Huenemoerder}, {Hamann},
  {Shenar}, {Sander}, {Ignace}, {Todt}, \& {Hainich}}]{2017ApJ...845...39O}
{Oskinova}, L.~M., {Huenemoerder}, D.~P., {Hamann}, W.~R., {et~al.} 2017, \apj,
  845, 39, \dodoi{10.3847/1538-4357/aa7e79}

\bibitem[{{Owocki} \& {Cohen}(2001)}]{Owocki:Cohen:2001}
{Owocki}, S.~P., \& {Cohen}, D.~H. 2001, \apj, 559, 1108,
  \dodoi{10.1086/322413}

\bibitem[{{Owocki} {et~al.}(2013){Owocki}, {Sundqvist}, {Cohen}, \&
  {Gayley}}]{2013MNRAS.429.3379O}
{Owocki}, S.~P., {Sundqvist}, J.~O., {Cohen}, D.~H., \& {Gayley}, K.~G. 2013,
  \mnras, 429, 3379, \dodoi{10.1093/mnras/sts599}

\bibitem[{{Pandey} {et~al.}(2014){Pandey}, {Pandey}, \&
  {Karmakar}}]{pandey:al:2014}
{Pandey}, J.~C., {Pandey}, S.~B., \& {Karmakar}, S. 2014, \apj, 788, 84,
  \dodoi{10.1088/0004-637X/788/1/84}

\bibitem[{{Parkin} \& {Gosset}(2011)}]{parkin:gosset:2011}
{Parkin}, E.~R., \& {Gosset}, E. 2011, \aap, 530, A119,
  \dodoi{10.1051/0004-6361/201016125}

\bibitem[{{Pittard}(2007)}]{2007ApJ...660L.141P}
{Pittard}, J.~M. 2007, \apjl, 660, L141, \dodoi{10.1086/518365}

\bibitem[{{Pittard} \& {Corcoran}(2002)}]{2002A&A...383..636P}
{Pittard}, J.~M., \& {Corcoran}, M.~F. 2002, \aap, 383, 636,
  \dodoi{10.1051/0004-6361:20020025}

\bibitem[{{Pittard} \& {Dawson}(2018)}]{pittard:dawson:2018}
{Pittard}, J.~M., \& {Dawson}, B. 2018, \mnras, 477, 5640,
  \dodoi{10.1093/mnras/sty1025}

\bibitem[{{Pittard} \& {Parkin}(2010)}]{2010MNRAS.403.1657P}
{Pittard}, J.~M., \& {Parkin}, E.~R. 2010, \mnras, 403, 1657,
  \dodoi{10.1111/j.1365-2966.2010.15776.x}

\bibitem[{{Pollock}(1987)}]{pollock1987}
{Pollock}, A.~M.~T. 1987, \apj, 320, 283, \dodoi{10.1086/165539}

\bibitem[{{Pollock}(2007)}]{pollock:2007}
---. 2007, \aap, 463, 1111, \dodoi{10.1051/0004-6361:20053838}

\bibitem[{{Pollock}(2012)}]{pollock:2012}
{Pollock}, A.~M.~T. 2012, in Astronomical Society of the Pacific Conference
  Series, Vol. 465, Proceedings of a Scientific Meeting in Honor of Anthony F.
  J. Moffat, ed. L.~{Drissen}, C.~{Robert}, N.~{St-Louis}, \& A.~F.~J.
  {Moffat}, 308

\bibitem[{{Pollock} \& {Corcoran}(2006)}]{pollock2006}
{Pollock}, A.~M.~T., \& {Corcoran}, M.~F. 2006, \aap, 445, 1093,
  \dodoi{10.1051/0004-6361:20053496}

\bibitem[{{Pollock} {et~al.}(2005){Pollock}, {Corcoran}, {Stevens}, \&
  {Williams}}]{pollock:al:2005}
{Pollock}, A.~M.~T., {Corcoran}, M.~F., {Stevens}, I.~R., \& {Williams}, P.~M.
  2005, \apj, 629, 482, \dodoi{10.1086/431193}

\bibitem[{{Pollock} {et~al.}(2018){Pollock}, {Crowther}, {Tehrani}, {Broos}, \&
  {Townsley}}]{PCTBT:2018}
{Pollock}, A.~M.~T., {Crowther}, P.~A., {Tehrani}, K., {Broos}, P.~S., \&
  {Townsley}, L.~K. 2018, \mnras, 474, 3228, \dodoi{10.1093/mnras/stx2879}

\bibitem[{{Prilutskii} \& {Usov}(1976)}]{1976AZh....53....6P}
{Prilutskii}, O.~F., \& {Usov}, V.~V. 1976, \azh, 53, 6

\bibitem[{{Puls} {et~al.}(2008){Puls}, {Vink}, \&
  {Najarro}}]{puls:vink:al:2008}
{Puls}, J., {Vink}, J.~S., \& {Najarro}, F. 2008, \aapr, 16, 209,
  \dodoi{10.1007/s00159-008-0015-8}

\bibitem[{{Rate} \& {Crowther}(2020)}]{rate:crowther:2020}
{Rate}, G., \& {Crowther}, P.~A. 2020, \mnras, 493, 1512,
  \dodoi{10.1093/mnras/stz3614}

\bibitem[{{Rauw} {et~al.}(2004){Rauw}, {De Becker}, {Naz{\'e}}, {Crowther},
  {Gosset}, {Sana}, {van der Hucht}, {Vreux}, \&
  {Williams}}]{rauw:debecker:al:2004}
{Rauw}, G., {De Becker}, M., {Naz{\'e}}, Y., {et~al.} 2004, \aap, 420, L9,
  \dodoi{10.1051/0004-6361:20040150}

\bibitem[{{Sander} {et~al.}(2017){Sander}, {Hamann}, {Todt}, {Hainich}, \&
  {Shenar}}]{2017A&A...603A..86S}
{Sander}, A.~A.~C., {Hamann}, W.~R., {Todt}, H., {Hainich}, R., \& {Shenar}, T.
  2017, \aap, 603, A86, \dodoi{10.1051/0004-6361/201730642}

\bibitem[{{Schild} {et~al.}(2004){Schild}, {G{\"u}del}, {Mewe}, {Schmutz},
  {Raassen}, {Audard}, {Dumm}, {van der Hucht}, {Leutenegger}, \&
  {Skinner}}]{cshild:al:2004}
{Schild}, H., {G{\"u}del}, M., {Mewe}, R., {et~al.} 2004, \aap, 422, 177,
  \dodoi{10.1051/0004-6361:20047035}

\bibitem[{{Schnurr} {et~al.}(2008){Schnurr}, {Casoli}, {Chen{\'e}}, {Moffat},
  \& {St-Louis}}]{schnurr:casoli:al:2008}
{Schnurr}, O., {Casoli}, J., {Chen{\'e}}, A.~N., {Moffat}, A.~F.~J., \&
  {St-Louis}, N. 2008, \mnras, 389, L38,
  \dodoi{10.1111/j.1745-3933.2008.00517.x}

\bibitem[{{Schulz} {et~al.}(2019){Schulz}, {Chakrabarty}, \&
  {Marshall}}]{schulz:al:2019pp}
{Schulz}, N.~S., {Chakrabarty}, D., \& {Marshall}, H.~L. 2019, arXiv e-prints,
  arXiv:1911.11684.
\newblock \doarXiv{1911.11684}

\bibitem[{{Schweickhardt} {et~al.}(1999){Schweickhardt}, {Schmutz}, {Stahl},
  {Szeifert}, \& {Wolf}}]{SSSSW:1999}
{Schweickhardt}, J., {Schmutz}, W., {Stahl}, O., {Szeifert}, T., \& {Wolf}, B.
  1999, \aap, 347, 127

\bibitem[{{Seward} {et~al.}(1979){Seward}, {Forman}, {Giacconi}, {Griffiths},
  {Harnden}, {Jones}, \& {Pye}}]{seward1979}
{Seward}, F.~D., {Forman}, W.~R., {Giacconi}, R., {et~al.} 1979, \apjl, 234,
  L55, \dodoi{10.1086/183108}

\bibitem[{{Skinner} {et~al.}(2001){Skinner}, {G{\"u}del}, {Schmutz}, \&
  {Stevens}}]{2001ApJ...558L.113S}
{Skinner}, S.~L., {G{\"u}del}, M., {Schmutz}, W., \& {Stevens}, I.~R. 2001,
  \apjl, 558, L113, \dodoi{10.1086/323567}

\bibitem[{{Smith}(2014)}]{smith:2014}
{Smith}, N. 2014, \araa, 52, 487, \dodoi{10.1146/annurev-astro-081913-040025}

\bibitem[{{Smith} {et~al.}(2001){Smith}, {Brickhouse}, {Liedahl}, \&
  {Raymond}}]{smith2001}
{Smith}, R.~K., {Brickhouse}, N.~S., {Liedahl}, D.~A., \& {Raymond}, J.~C.
  2001, \apjl, 556, L91, \dodoi{10.1086/322992}

\bibitem[{Sota {et~al.}(2014)Sota, Apell{\'{a}}niz, Morrell, Barb{\'{a}},
  Walborn, Gamen, Arias, \& Alfaro}]{Sota_2014}
Sota, A., Apell{\'{a}}niz, J.~M., Morrell, N.~I., {et~al.} 2014, The
  Astrophysical Journal Supplement Series, 211, 10,
  \dodoi{10.1088/0067-0049/211/1/10}

\bibitem[{{Stevens} {et~al.}(1992){Stevens}, {Blondin}, \&
  {Pollock}}]{Stevens:Blondin:al:1992}
{Stevens}, I.~R., {Blondin}, J.~M., \& {Pollock}, A.~M.~T. 1992, \apj, 386,
  265, \dodoi{10.1086/171013}

\bibitem[{{Sundqvist} {et~al.}(2012){Sundqvist}, {Owocki}, {Cohen},
  {Leutenegger}, \& {Townsend}}]{2012MNRAS.420.1553S}
{Sundqvist}, J.~O., {Owocki}, S.~P., {Cohen}, D.~H., {Leutenegger}, M.~A., \&
  {Townsend}, R. H.~D. 2012, \mnras, 420, 1553,
  \dodoi{10.1111/j.1365-2966.2011.20141.x}

\bibitem[{{Sundqvist} {et~al.}(2018){Sundqvist}, {Owocki}, \&
  {Puls}}]{2018A&A...611A..17S}
{Sundqvist}, J.~O., {Owocki}, S.~P., \& {Puls}, J. 2018, \aap, 611, A17,
  \dodoi{10.1051/0004-6361/201731718}

\bibitem[{{Tehrani} {et~al.}(2019){Tehrani}, {Crowther}, {Bestenlehner},
  {Littlefair}, {Pollock}, {Parker}, \& {Schnurr}}]{TCBLPPS:2019}
{Tehrani}, K.~A., {Crowther}, P.~A., {Bestenlehner}, J.~M., {et~al.} 2019,
  \mnras, 484, 2692, \dodoi{10.1093/mnras/stz147}

\bibitem[{{Tramper} {et~al.}(2016){Tramper}, {Sana}, {Fitzsimons}, {de Koter},
  {Kaper}, {Mahy}, \& {Moffat}}]{tramper:sana:al:2016}
{Tramper}, F., {Sana}, H., {Fitzsimons}, N.~E., {et~al.} 2016, \mnras, 455,
  1275, \dodoi{10.1093/mnras/stv2373}

\bibitem[{{Usov}(1992)}]{1992ApJ...389..635U}
{Usov}, V.~V. 1992, \apj, 389, 635, \dodoi{10.1086/171236}

\bibitem[{{Walborn} {et~al.}(2009){Walborn}, {Nichols}, \&
  {Waldron}}]{Walborn:Nichols:Waldron:2009}
{Walborn}, N.~R., {Nichols}, J.~S., \& {Waldron}, W.~L. 2009, \apj, 703, 633,
  \dodoi{10.1088/0004-637X/703/1/633}

\bibitem[{{Waldron} \& {Cassinelli}(2001)}]{Waldron:Cassinelli:2001}
{Waldron}, W.~L., \& {Cassinelli}, J.~P. 2001, \apjl, 548, L45,
  \dodoi{10.1086/318926}

\bibitem[{{Waldron} \& {Cassinelli}(2007)}]{Waldron:Cassinelli:2007}
---. 2007, \apj, 668, 456, \dodoi{10.1086/520919}

\bibitem[{{Williams} {et~al.}(1990){Williams}, {van der Hucht}, {Pollock},
  {Florkowski}, {van der Woerd}, \& {Wamsteker}}]{williams:al:1990}
{Williams}, P.~M., {van der Hucht}, K.~A., {Pollock}, A.~M.~T., {et~al.} 1990,
  \mnras, 243, 662

\end{thebibliography}

\end{document}